\documentclass[10pt,a4paper]{article}
\pdfoutput=1
\usepackage{amsmath}
\usepackage{amsthm}
\numberwithin{equation}{section}
\usepackage{graphicx,lscape}
\title{Cross Ranking of Cities and Regions: Population vs. Income}
\author{Roy Cerqueti$^{1,}$\footnote{Corresponding address: University of Macerata, Department of Economics and Law,   via Crescimbeni 20, I-62100, Macerata, Italy. Tel.: +39 0733 258 3246; Fax: +39 0733 258 3205. Email: roy.cerqueti@unimc.it} $\;$ and  Marcel Ausloos$^{2,3}$}

 \date{
$^1$ University of Macerata, Department of Economics and Law,\\  via
Crescimbeni 20,   I-62100, Macerata, Italy \\  $e$-$mail$ $address$:
roy.cerqueti@unimc.it\\  \vskip0.5cm $^{2}$eHumanities
group\footnote{Associate Researcher}$\;$, \\Royal Netherlands
Academy of Arts and Sciences, \\  Joan Muyskenweg 25, 1096 CJ
Amsterdam, The Netherlands \\ \vskip0.5cm $^3$GRAPES\footnote{Group
of Researchers for Applications of Physics in Economy and Sociology
}$\;$,
\\   rue de la Belle Jardiniere 483, B-4031, Angleur Liege, Belgium \\$e$-$mail$ $address$:
marcel.ausloos@ulg.ac.be
 }
\begin{document}
 \maketitle
\begin{abstract}
This paper explores the relationship between the inner economical structure  of communities and  their population distribution through a
rank-rank analysis of official data, along statistical physics ideas  within two techniques.  The data is taken on Italian cities. The analysis is performed both at a global (national)  and at a  more local  (regional)  level in order to distinguish "macro" and "micro" aspects.   First, the rank-size rule is  found not to be a standard power law, as in many other studies, but a doubly decreasing power law.  Next,  the Kendall $\tau$  and the Spearman $\rho$ rank correlation coefficients which measure pair concordance and the correlation between fluctuations in two rankings, respectively,  - as a correlation function does  in thermodynamics,    are calculated for finding rank correlation (if any) between demography and wealth.  Results show   non only global disparities for the whole (country) set,  but also (regional) disparities, when comparing the number of cities in regions, the number of inhabitants in cities and  that  in regions, as well as when comparing the aggregated tax income of the cities and  that of regions. Different outliers are pointed out and justified. Interestingly,  two classes of cities in the country and two classes of  regions in the  country are found. "Common sense" social, political, and economic considerations sustain the findings. More importantly, the methods   show that they allow  to distinguish  communities,  very clearly, when specific criteria are numerically sound. A specific modeling  for the findings is presented, i.e. for the doubly decreasing power law and the two phase system, based on  statistics theory, e.g., urn filling. The model    ideas can be expected to hold when similar rank relationship features are observed in fields. It is emphasized that the analysis makes more sense  than one through a Pearson $\Pi$ value-value correlation analysis.
\end{abstract}

\section{ Introduction}\label{Introduction}

Research on rank-size relationships has a long history and has been
applied in a wide range of contexts. In this respect, at the
inception Zipf's law \cite{zipf} was illustrated through linguistics
considerations, while Pareto's law -- a similar hyperbolic power law
-- finds its origin in finance \cite{newman}.
\newline
However, among several applications, rank-size theory has   a
prominence in the field of urban studies
\cite{zipf,jefferson,beckmann,gabaix,bettencourt,Semboloni}.
 In particular, a relevant role is played by the analysis of the geographical-economical variables in the conceptualization of the
New Economic Geography, introduced by Krugman \cite{krugman}. In this respect, see Berry  \cite{Berry}, Pianegonda and Iglesias \cite{Pianegonda} and the extensive surveys of Ottaviano and Puga \cite{ottp}, Fujita et al. \cite{fkv}, Neary \cite{neary}, Baldwin et al. \cite{baldwin},  and Fujita and Mori \cite{fm}.
 Such an analysis can be satisfactorily developed through the study of the rank-size rule for regional and urban areas, on the
basis of economical variables and the population size distribution, as shown here below. The interested reader is referred to the monograph of Chakrabarti et al. \cite{Chakrabarti} for outlining interdisciplinary socio-econo-physics points of view.

Several studies proved empirically the validity of Zipf's law \cite{zipf} (or type-I Pareto distribution \cite{BGNEGNKVZID}): Rosen and Resnick \cite{RosenResnick}, in 1980,
analyzed data from 44 Countries, and found a clear predominance of statistical significance of Zipf's law, with $R^2$ greater than 0.9 (except in one case, Thailand);
 in Mills and Hamilton  \cite{MillsHamilton},
 data from US city sizes, in 1990, has been taken to show the evidence of Zipf's law ($R^2\sim 0.99$); see  Guerin-Pace \cite{Guerin-Pace} also. 
 Other papers, after 2000,  which substantially support this type of rank-size rule are
Dobkins and Ioannides \cite{DobkinsIoannides}, 
Song and Zhang \cite{SongZhang}, 
Ioannides and Overman  \cite{IoannidesOverman}, 
Gabaix and Ioannides \cite{GabaixIoannides04} 
Reed \cite{Reed},  
and Dimitrova and Ausloos  \cite{ZDMA} more recently, but with warnings \footnote{In a practically modeling approach, Dimitrova and Ausloos   \cite{ZDMA}  indicated  through the notion of the global
primacy index of Sheppard   \cite{s1} that Gibrat  (growth)
law \cite{Gibrat}, supposedly at the origin of Zipf's law, in fact, does not hold in the case of Bulgaria cities, but can be valid when selecting various city classes (large  or small sizes).},  
 just to cite a few.   Nitsch
 \cite{Nitsch} 
provides an exhaustive literature review up to 2005.

As other counter-examples,   beside the above-mentioned case of Thailand in \cite{RosenResnick}  and Bulgaria in  \cite{ZDMA}, weak  agreement  between data and Pareto fit is sometimes pointed out.   Peng  \cite{Peng} 
  found a Pareto coefficient of $0.84$, not  quite close to 1, - when implementing a best fit of data
on Chinese city sizes in 1999-2004  with the Pareto distribution. Ioannides and Skouras
 \cite{IoannidesSkouras}, 
among others, have argued that Pareto-Zipf's law seems to stand in force only in the tail of the data distribution. Matlaba et al. \cite{Matlabaetal} 
also provided  such an "evidence that, at least for the analyzed case of Brazilian urban areas over a spectacularly wide period (1907-2008), Zipf's law is clearly rejected.  Soo  \cite{Soo} 
has also empirically  shown   that the size of Malaysian cities cannot be
plotted according to such a rank-size rule, but a suitable collection
of them can do it.


A list of other contributions on the inconsistency of Zipf's law in
several countries, different periods and under specific economic
conditions should include \cite{Cordoba}, \cite{Garmestanietal2007}
and \cite{Boskeretal2008}. Of particular interest, in the present
case,  is also Garmestani et al. \cite{Garmestanietal2008}, who
conducted an analysis for the USA at a regional level. Thus, it
seems that the failure of Zipf's law may often depend  on the way
data are grouped  \cite{ZDMA,GiesSud}

From the present state of the art point of view, regional
agglomerations,  commonly ranked in terms of population, may be also
sorted out in an order dealing with the economic variables. In fact,
Zipf's law is sometimes  identified   in some ''economic'' way to
rank. As an example, Skipper  \cite{Skipper011} used such  a
rank-size relationship to  detect   well developed countries
hierarchy  through their national GDP. This result has been also
achieved by Cristelli et al.  \cite{Cristellietal12}, who exhibited
evidence of the Zipf's law for the top fifty richest countries in
the 1900-2008 time interval.

{\it In fine}, the investigation, in the main text, aims to provide
some newness,  through some recent data; even leading to a better
description of a rank-size rule  than the Pareto-Zipf's law. Beside,
to our knowledge,  no statistical evidence of Zipf's law studies has
been reported  for the economic variables characterizing Italian
cities and regions, in the period 2007-2011.

Along such lines of thought and within our statistical physics framework,  the paper deals with the rank-size rule for the entire set of
municipalities in Italy (IT, hereafter). This country, Italy, is expected, according to its fame,   to provide what a physicist looks for, i.e. some universality features but also some non universal  ones.  Therefore,   in an aim toward understanding nature and progressing  toward reconciling so called hard and soft science,   reliable data  is investigated 
looking for "universality" and "deviations".  The IT data are both
official, and are given by aggregated income tax (ATI) and number of
inhabitants : the former has been provided directly from the
Research Center of the Italian Ministry of Economics and Finance
(MEF), and cover the quinquennium 2007-2011; the latter comes from
the Census 2011, which has been performed by the Italian Institute
of Statistics (ISTAT).

Therefore, the \textit{size} to be examined is here defined through
two criteria: (i) by the ATI  \textit{contribution} that each city
has given to the Italian GDP and (ii) by the \textit{population} of
each city. First, the 8092 Italian cities are yearly ranked
according to such variables. Their related classifications are then
compared: (i)  at the national  level,  but also (ii)  at the
regional level.  There are 20 regions in IT with a varied number of
cities.

Each specific year  of the quinquennium has been examined. However,
special attention has been paid to 2011, in order to be somewhat
consistent with the year concerned by the Italian census report on
population. The census took several years in fact. Therefore, only
the ATI averaged over the 5 years interval for each municipality is
reported in the main text and discussed through  the mean  value
over 5 years of the yearly ATI data.  The conclusions are
unaffected,   as discussed in an Appendix, except for some mild
change in error bars on the numerical parameters, when specific
years are selected for examination.

Within the statistical physics approach interested in correlation
functions, the paper  also aims at observing whether there is or not
some correlation between ATI and population rankings.  For  this
aim, the Kendall $\tau$ and the Spearman $\rho$ rank correlation
measures have been computed
\cite{kendall,Abdi,Spearman,BolboacaaJantschi}. The Kendall  $\tau$
measure compares the number of concordant and non-concordant pairs,
to detect the presence of singularities in a possible relationship
(here between economy and demography). The Spearman $\rho$ rank
correlation is also computed, to provide a more satisfactory
interpretation of the relationship between economy and demography in
terms of rank \cite{Bland}.

A rank scatter plot  of  the number of inhabitants in each 8092 city
versus the corresponding ATI reveals two regimes. Moreover, two
regimes are also observed for the value themselves, distinguishing
different "phase states", pointing to regional  specificities. It
can be stressed \cite{Abdi}  that such a rank-rank analysis,
implying pair correlations, is the analog of a correlation function,
sometimes called susceptibility, in the linear response theory of
statistical mechanics. It will be indicated that a "rank" is like a
"temperature".  A useful methodological paper  to read on the
rank-rank correlation is  by Melucci  \cite{melucci}. It contains
also a bibliographic review on previous researches on this theme, up
to that time.

In order to obtain analytical expressions simulating the data,
whence suggesting a model susceptible of general applications,
various simple fits have been attempted. The most classical one, for
universality purpose, is the straight line fit on a log-log plot.
However, both for the population size and the ATI data, it turns out
that, in each region, the main city is an outlier\footnote{This was
observed  already by   Jefferson \cite{jefferson}.}:  more precisely
25 cities for the whole country, i.e. about 1 per region. These
lowest rank cities are markedly  found to occur much above the usual
expecting straight line data fit on a log-log plot (the 20
"regional" plots are not shown for saving space). Moreover, the fit
visual appearances are not exciting, because our eyes receive the
same effects from  the low and high  rank ranges. Practically, it
has been found that the regression  coefficient $R^2$ improves if
one removes these outliers.  Moreover,  the fits visually improve in
the asymptotic regimes, - which are very narrow regions, in
particular in the high rank range. Therefore, for shortening the
paper, the  parameters of the fits reported here below only concern
fits with a 3-parameter function, discussed below.

To get some perspective, notice that several  contributions in the
literature propose  rank-rank analysis  types within different
contexts, -  all comparing two different rank rules.  In a series of
papers \cite{ausloos2014b,ausloosetal,ausloosetal2014b},
country ranking due  to soccer team ranking (and performance) in
UEFA  competitions is presented and contrasted with FIFA ranking. In
particular, in dealing with the rank-rank correlation
\cite{ausloosetal},  the Kendall $\tau$ is employed. Interesting is
also the application in the context of archeology, with a specific
focus on the Aztec settlement distributions, presented in Hare
\cite{hare}. In \cite{njp}, Zhou et al.  focus on the rank-rank
correlation for scientists and scientific journal, in line with the
scientometrics literature. In this respect, see also
\cite{ausloos2013},
 - on the relationship between authors and coauthors,
and Stallings et al. \cite{pnas}, - comparing researchers and
universities according to different criteria, whence  providing also
an axiomatization of the rank and rank correlation problems. 
For what concerns the Spearman $\rho$ coefficient, it measures
correlations in the rank deviation from their mean  of the
measurements. As already said above, for completeness,  $\rho$ will
be calculated, even if Kendall $\tau$ is usually acknowledged to
have better statistical properties than Spearman's $\rho$
\cite{Bland,Hsieh2010,Yangetal2010,Schmitzetal2013,Rezapour2013}.
The interpretation of Kendall $\tau$  also seems to be easier and
more intuitive than that of the Spearman $\rho$. In this respect, it
is important to point out that the average of the ranks is equal to
$N/2$, thus representing a measure of the sample size $N$.

It is  of common knowledge that the Pearson  $\Pi$ correlation
coefficient is  the most
 frequently used measure when data is supposed to be (or is) normally
 distributed. On the other hand, nonparametric methods such as
 Spearman's rank-order and Kendall's correlation coefficients are
 usually suggested  to be better for non-normal data analysis.  In
 fact, all three correlation coefficients are proportional to
differently weighted averages of the concordance indicators.  The
normality  constraint is  briefly examined in  an Appendix, together
with a discussion of  the Pearson coefficient interest  in the
present context.  It is briefly argued  that the Kendall $\tau$
brings some information of  interest. We consider that   this is
particularly  so,  when values   are of different natures  and are
measured with a quite  often unknown error bar.

In the field of Economic Geography, the rank-rank analysis is quite neglected. A noteworthy exception is Rappoport  \cite{rapp}, where population densities and consumption amenities are compared and
discussed for U.S. economical-demographical data. The paper of Mori and Smith  \cite{mori} is also of interest, in that it focuses on the link between economics and demography at a city level by investigating
the number of cities inhabitants in presence of established industries. However, to the best of our knowledge, the main text below is the first paper dealing with the application of this theory for discussing the relationship between (Italian) demographical and economical reality.

It is also important to note that the employment of microdata allows
to emphasize Italian regional disparities. In this respect, we
address the interested reader to De Groot et al. \cite{degroot2009} and Melo et
al.  \cite{melo}.

In short, the paper is organized as follows: Section \ref{8092}
contains the description of the data. Section
\ref{citydistributions} is devoted to the analysis of the whole IT.
Two measures,  the Kendall's $\tau$  and the  Spearman's $\rho$ rank
correlation coefficients,  are proposed and their respective
interest discussed.  The findings are   collected and  reworded in
Section \ref{sec:results}. Such section includes also Subsection
\ref{regionaldisparities}, which serves to emphasize the
regularities and disparities between   the  Italian regions.
Thereafter, a specific modeling based on  urn filling statistics,
 is presented in Section \ref{Model}. It can be expected to hold when
 rank relationship features similar to those of our findings  are observed. The last section
(Section \ref{conclusions})  serves for conclusions and for offering
suggestions for further research lines.

There are three  Appendices:  App. A contains  a technical detail note on data aggregation, as already mentioned, arising from the change in the number of cities  in Italy during the ATI measurements.   
 App. B is  a  short investigation  on the (as also pointed here above) negligible, in fact,  time dependence.
App. C  contains a note on the Pearson coefficient and some argument
in favor of  considering  rank-rank correlations instead of
value-value correlations.

\section{Data}\label{8092}
The population data comes from the  2011 Census performed by the
Italian Institute of Statistics (ISTAT)\footnote{Census is an
official statistical exploration of the Italian population. It is
based on the responses provided by all the Italians, and it is
performed every 10 years. However, there were Irregularities: in 1891 and 1941 Census
has been not performed (for financial distress in the former case, and  due to
the Second World War in the latter one), but an adjunctive Census
has been provided in 1936. The next Census will be in 2021.}.
 The economic data 
has been obtained by and from  the Research Center of the Italian
Ministry of Economics and Finance.
Population data consist of number of inhabitants, while economic
data are given by IT GDP for recent five years (2007-2011) (the
aggregated tax income - ATI). In both cases, we have disaggregated
contributions at a municipal level (in IT a \textit{municipality} or
\textit{city} is denoted as \textit{comune}, - plural $comuni$).

To provide some better understanding of the paper aims and results,
the IT administrative structure is here first described.

 IT is composed of 20 regions, more
than 100 provinces and more than 8000 municipalities. Each municipality belongs to one
and only one province;  each province is contained in one and only one region. Administrative modifications  due to the IT political system has led to a varying number of provinces and
municipalities during the examined  quinquennium. The number of
cities in each  administrative entity has also changed, but the number of regions has been constantly
equal to 20.  

 Therefore, the  available yearly ATI data
corresponds to a different number of cities. In particular, the
number of cities has been yearly evolving respectively  as follows :
8101, 8094, 8094, 8092, 8092,  -  from 2007 till 2011. Details are given in Appendix A.

The  number of cities in  each region  is  given as a function of
time in Table \ref{TableNcityperregion}. \newline For making sense,
it is necessary to compare identical lists.  We have considered this
latest 2011 "count" as the basic one.  Therefore, we have taken into
account a virtual merging of cities, in the appropriate  (previous
to 2011) years, according to IT administrative law statements (see
also $http://www.comuni-italiani.it/regioni.html$). 

 In the same spirit, the ATI
of the resulting cities  (and regions) have been
linearly adapted, as if these ATI were  existing before the merging or city phagocytosis.

A summary of   the statistical characteristics for the ATI of  all
($r_M \equiv N=8092$) IT cities in 2007-2011 is  reported in Table
\ref{Tablestat}.
For a statistical  overview of the Italian structure, at the
regional level, see Table \ref{Tablestatcityperregion}.

It is also worth noting that there is some change in the ATI rank of a city as times goes by. 
Care was taken that the arithmetics pertain to the same city,  when
a sum or average was made. For example, there are twice 3 cities with the
same name in IT; we carefully distinguished them.

\begin{table} \begin{center}
\begin{tabular}[t]{ccccc}
  \hline
   region&&$N_{c,r}$ \\ \hline
  year: &2007&2009&2011 \\ \hline
 Lombardia& 1546     &1546  &   $\downarrow$    1544\\
Piemonte    &1206\\
Veneto  &581\\
Campania&   551\\
Calabria&   409\\
Sicilia&    390\\
Lazio   &378\\
Sardegna    &377\\
Emilia-Romagna&  341&341    & $\uparrow 348$ \\
Trentino-Alto Adige&  339   &$\downarrow$   333&333\\
Abruzzo&    305\\
Toscana &287\\
Puglia& 258\\
Marche  &246&246    &$\downarrow$   239\\
Liguria&    235\\
Friuli-Venezia Giulia& 219  &$\downarrow$   218 &218\\
Molise  &136\\
Basilicata  &131\\
Umbria& 92\\
Valle d'Aosta&  74\\  \hline
Total &8101&$\downarrow$    8094& $\downarrow$  8092\\\hline
\end{tabular}
\caption{Number $N$ of (8092 $\equiv r_M$) cities in 2011, and in
previous years, in  the (20) IT regions;  such a region ranking by
city number corresponds to that illustrated in Fig. \ref{Plot20N20}.
} \label{TableNcityperregion}
\end{center} \end{table}

\begin{table} \begin{center}
\begin{tabular}[t]{cccccccc}
  \hline
   $ $   &2007 &2008 &2009&2010&   2011 &5yr av. \\
\hline
min. (x$10^{-5}$)   &3.0455         &2.9914      &   3.0909    &3.6083        &3.3479& 3.3219  \\
Max. (x$10^{-10}$)&   4.3590&4.4360 &    4.4777      &4.5413 &4.5490&4.4726 \\
Sum (x$10^{-11}$)&6.8947 &7.0427 &    7.0600 &7.1426 &7.2184  &7.0738    \\
 Max. range ($r_M$)  &8092      &8092       &   8092    &8092       &8092 &8092   \\
mean ($\mu$) (x$10^{-7}$)   &8.5204  &8.7033     &   8.7248&8.8267 &8.9204 &8.7417 \\
median ($m$) (x$10^{-7}$)  &2.2875 &2.3553 &   2.3777 &2.4055&2.4601 &2.3828 \\
RMS (x$10^{-8}$)    &6.5629 &6.6598 &    6.6640&6.7531 &6.7701&6.682 \\
Std. Dev. ($\sigma$) (x$10^{-8}$) &6.5078&6.6031&   6.6070& 6.6956 &6.7115 &6.6256\\
Var.    (x$10^{-17}$)&4.2351&4.3601&    4.3653 &4.4831 &4.5044&4.3899 \\
Std. Err. (x$10^{-6}$)&7.2344 &7.3404  &   7.3448&7.4432 &7.4609&7.3654 \\
Skewness  &48.685 &48.855&   49.266&49.414 &49.490&49.126 \\
Kurtosis      &2898.7     &2920.42   &   2978.1       &2991.0 &2994.7  &2955.2    \\  \hline
 $\mu/\sigma$  &0.1309   &0.1318&0.1321 &0.1319&  0.1329& 0.1319  \\
$3(\mu-m)/\sigma$  &0.2873&0.2884&0.2883 &0.2878& 0.2889& 0.2879   \\
 \hline
 \end{tabular}
   \caption{Summary of  (rounded) statistical characteristics  for ATI (in EUR) of IT cities ($N=8092$) in   2007-2011.}\label{Tablestat}
 \end{center} \end{table}

\begin{table} \begin{center}
\begin{tabular}[t]{cccc}
\hline &
&$N_{c,r}$\\
  \hline
Minimum &   
&   74  \\
Maximum &  
&   1544    \\
Mean     ($\mu$) &  
&   404.6   \\
Median  ($m$)&  
&   319 \\
RMS &   
 &   536.998     \\
Std Dev     ($\sigma$)&   
&   362.253     \\
Variance    &   
&   131 227.52  \\
Std Error   &  
&   81.0023     \\
Skewness    &   
&    2.1284\\
Kurtosis    &  
&    3.8693\\
  \hline
 $\mu/\sigma$ & 
  & 1.117 &\\
$3(\mu-m)/\sigma$ & 
&0.7089 &   \\
\hline
\end{tabular}
\caption{Summary of  (rounded) statistical characteristics for the
number ($N_c=8092$) distribution of IT cities in the various
regions   ($N_r=20$) in 2011.
 The maximum
$N_{c,r}$ is 1544  (Lombardia) and the minimum is 74  (Valle
d'Aosta), - see Table \ref{TableNcityperregion}.}
\label{Tablestatcityperregion}
\end{center} \end{table}


 \begin{figure}
\centering
\includegraphics[height=10.0cm,width=12.7cm] {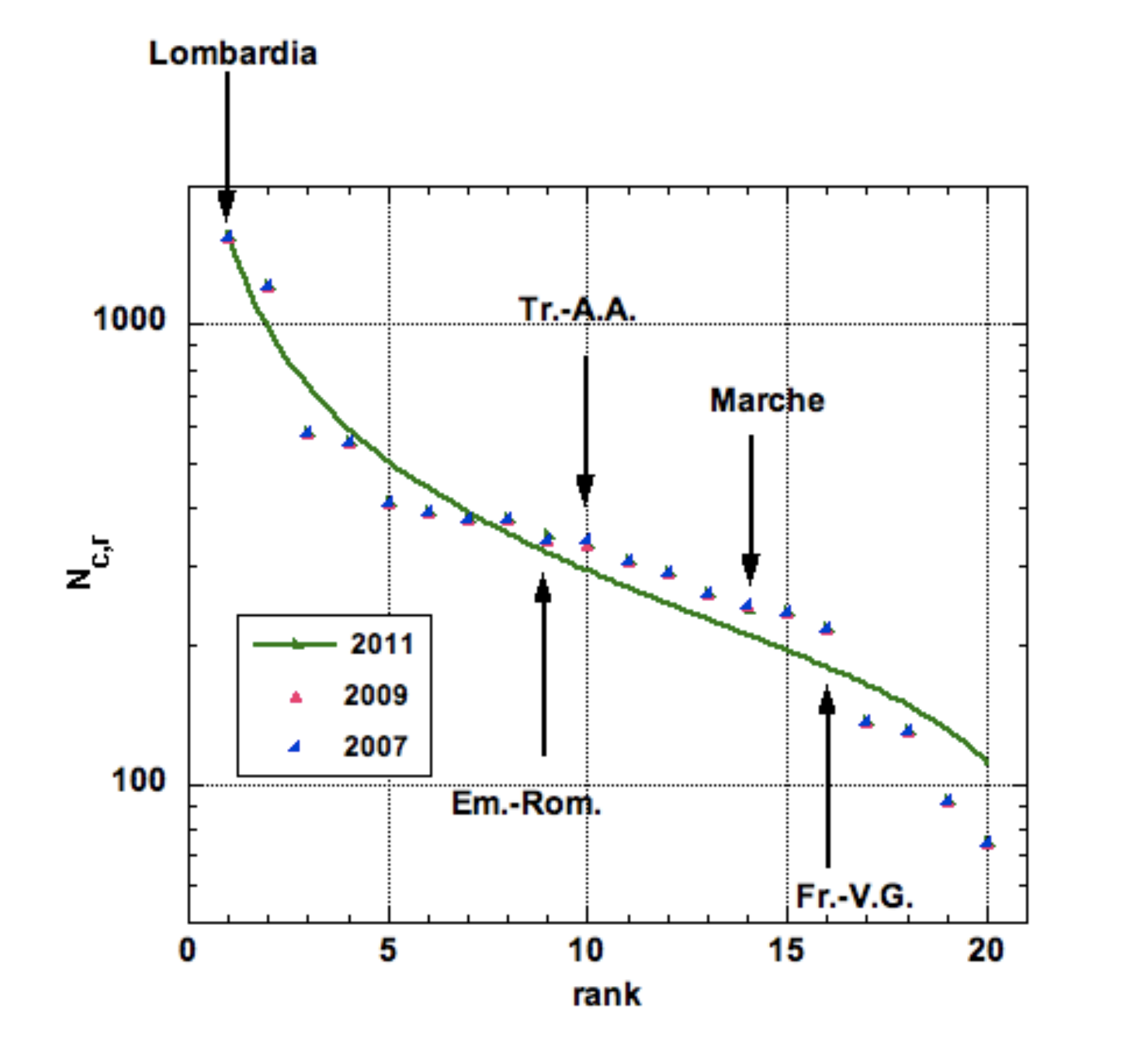}
 \caption{ $N_{c,r}$ $vs.$ rank of the region for the years of the quinquennium;
 the regions having a change in the number of cities are indicated by an arrow $\uparrow$ or $\downarrow$;     the arrow direction is according to the change in $N_{c,r}$ in some year as mentioned in Table \ref{TableNcityperregion}. The fit corresponds to the function    Eq. (\ref{Lav3});  the fit parameters are given in the text.} \label{Plot20N20}
\end{figure}

  \begin{figure}
\centering
\includegraphics[height=10.0cm,width=12.7cm]
{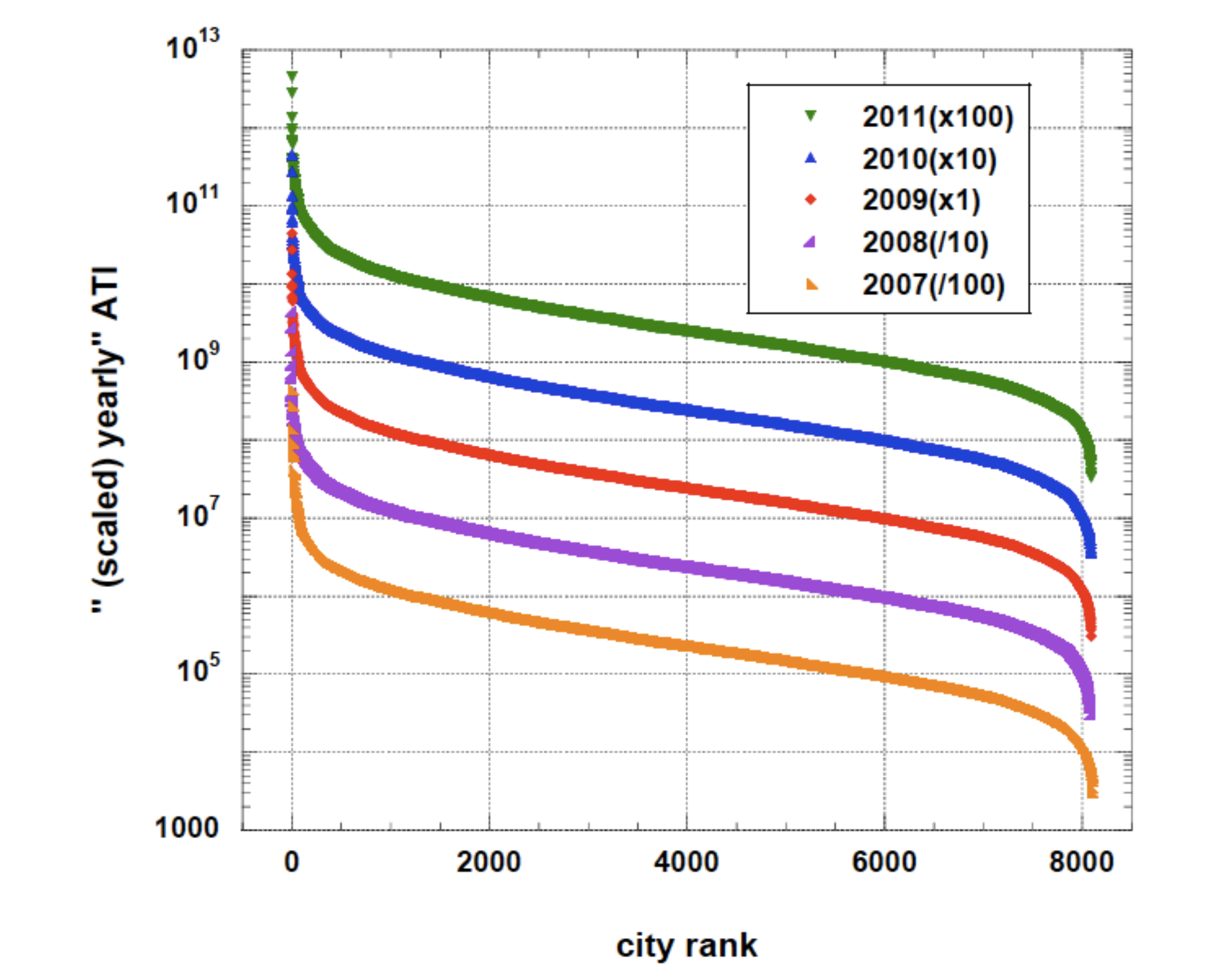} \caption   { Semi-log plot  of the
2007-2011 yearly ATI of the  8902 IT cities ranked according to
their "income tax" importance every year; the data is rescaled by a
factor 10 or 100, as indicated in the insert,  for better
visibility. The inflection point  is well seen near $r_M/2 \sim
4000$. }
 \label{fig:Plot35plots5yliloLavlike}
\end{figure}

  \begin{figure}
\centering
\includegraphics[height=10.0cm,width=12.7cm] {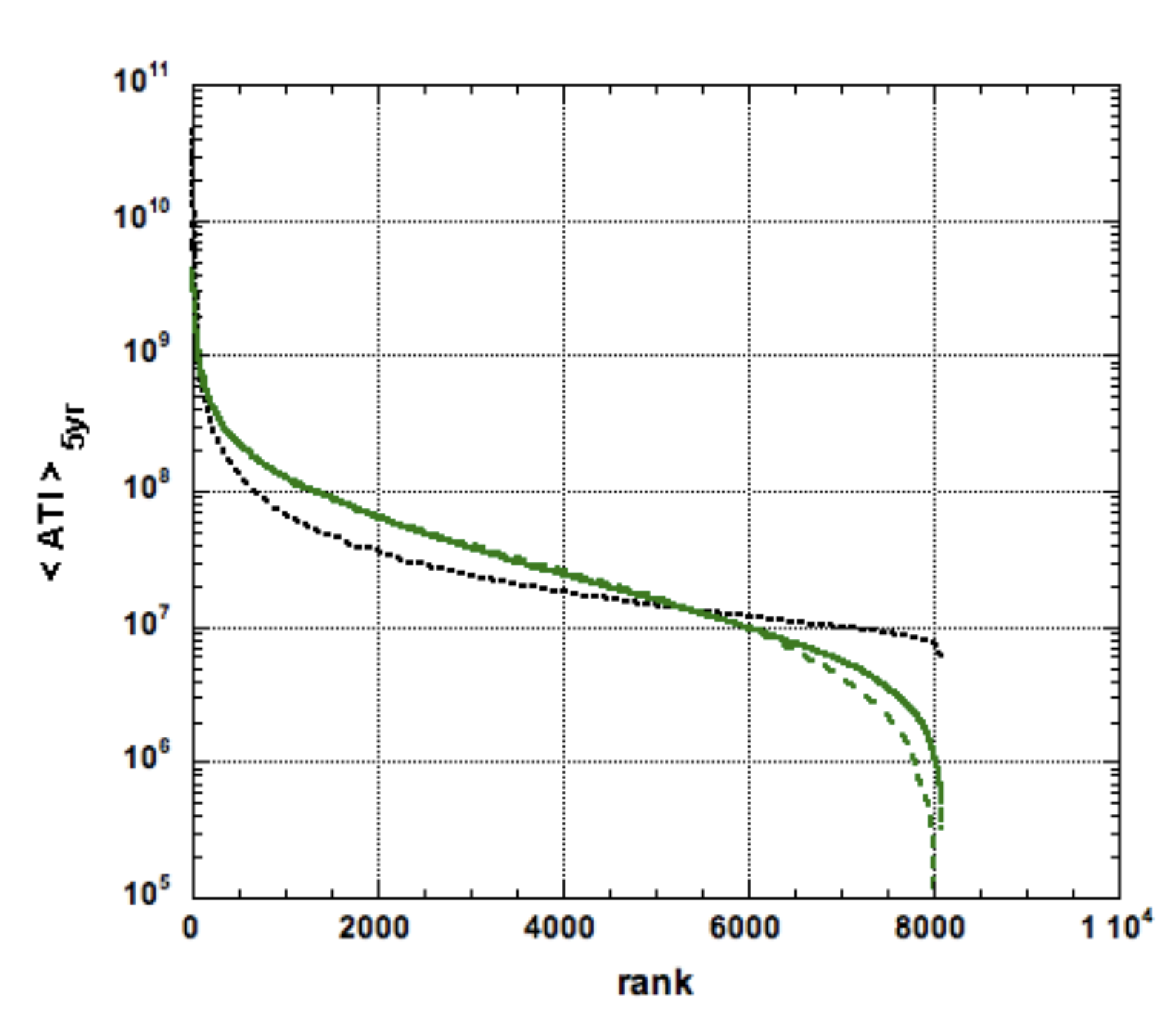}
 \caption{
Semi-log plot of the rank-size  relationship between  each  Italian
city   $<ATI>$  (averaged for the  examined quinquenium)  and  its
rank;  the black dot line corresponds to the whole (8092) data; the
green dash line corresponds to  the  whole data minus the top 8 city
outliers.} \label{Plot21ATI5yrav2fits}
\end{figure}

  \begin{figure}
\centering
 \includegraphics[height=10.0cm,width=12.7cm]
{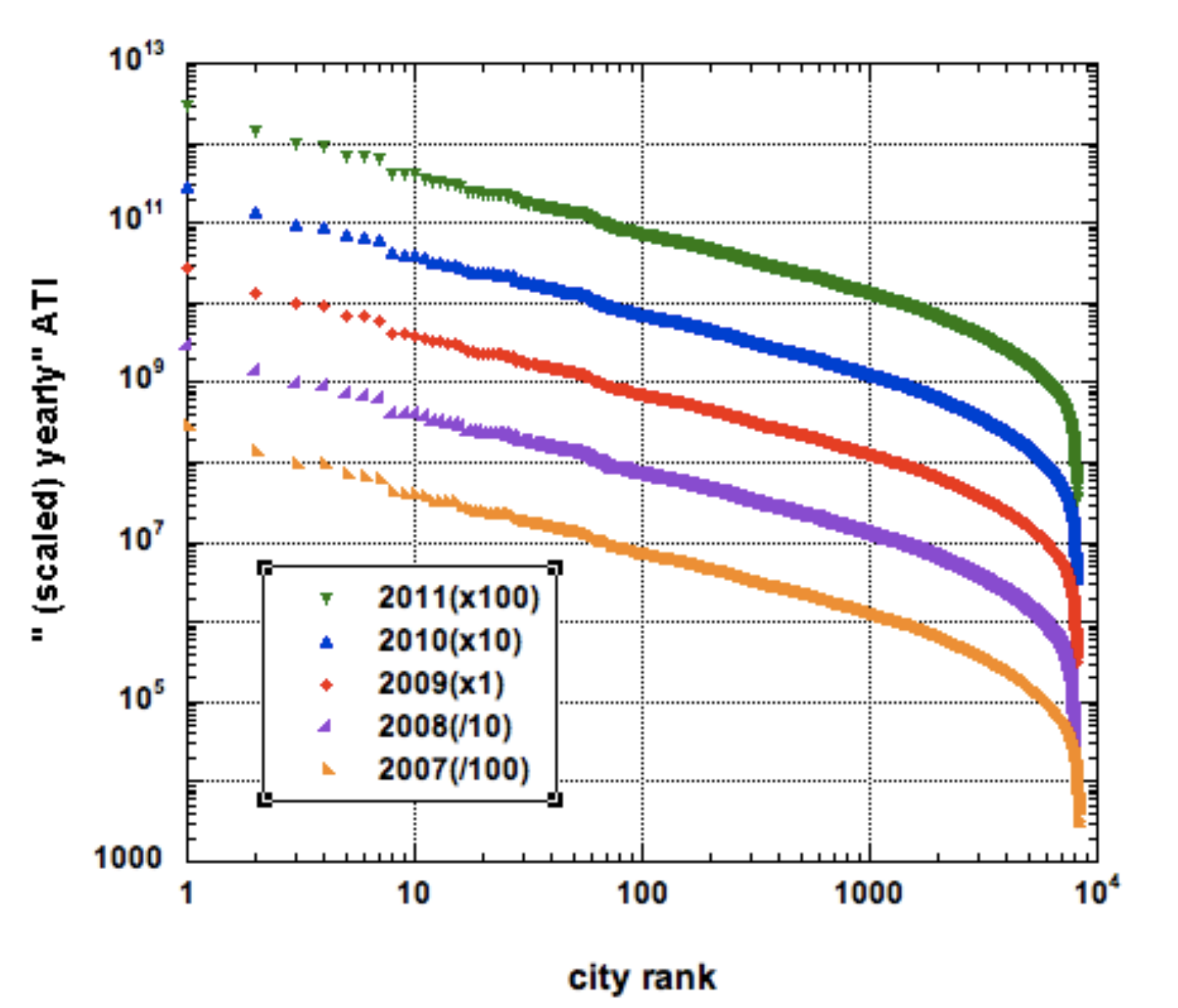} \caption   { Log-log plot  of the 2007-2011 yearly
ATI of the  8902 IT cities ranked according to their "income tax"
importance every year; the data is rescaled by a factor 10 or 100,
as indicated in the insert,  for better visibility. The outliers are
here better emphasized than on Fig.
\ref{fig:Plot35plots5yliloLavlike} but the inflection point  near
$r_M/2 \sim 4000$ not so obvious.}
 \label{Plot35plotslolo}
\end{figure}

\begin{table} \begin{center}
\begin{tabular}[t]{cccc}
  \hline
&(i) (8092) &  (ii) (20)\\ \hline
$p+q$&32 736 186 &190 \\
 $  p-q$    &27 778 116  &148\\
$p$&30 256 042 & 169 \\
$q$&2 480 144 & 21 \\
Kendall $\tau$& 0.849   &0.779 \\
 \hline
Spearman $\rho$& 0.9637   & 0.9098 \\
 \hline
Pearson $\Pi$& 0.9849   & 0.9787 \\
 \hline
\end{tabular}
\caption{Kendall $\tau$, Eq. (\ref{taueq}) and Spearman $\rho$, Eq.
(\ref{Spearmanrho}) correlation statistics of ranking order  between
the Number of inhabitants, (i)  in  (8092) cities  or (ii) in (20)
regions, according to the 2011 Census,  and the  corresponding
averaged ATI, over the period $2007-2011$;  the  Pearson $\Pi$
value-value correlation coefficient is given for completeness.
 } \label{Tabletaurank}
\end{center} \end{table}

\section{City population size and ATI rank order     distributions in IT 
 } \label{citydistributions}
\vskip 0.5truecm
In this section, $size$ is defined under two criteria:  either (i)
the economic one (averaged ATI over the period
2007-2011)\footnote{In the Appendix B it is verified that taking the
average of the ATI over the considered quinquennium does not bias
the analysis.}  or (ii)  the demographic one (population in 2011).
The  empirical rank-size relationships are first looked for, see
Sect. \ref{Rank-size relationships}. The Kendall $\tau$ coefficient
is next used to compare rank pairing  under both criteria, in Sect.
\ref {Kendall coefficient}. The Spearman $\rho$ coefficient is next
calculated and  compared under both criteria, in Sect. \ref {Kendall
coefficient}. The Pearson $\Pi$ coefficient is discussed in App. C.

\subsection{Rank-size relationships}\label{Rank-size relationships}

We have ranked the  regions in decreasing order, according to their
respective number of cities. In general,   the central part of the
data looks like being well fitted by a power law with an exponent a
little bit below (-1). However, at low rank, there is usually a
jump, while   at high rank,  there is a  sharply marked downward curvature.

Therefore, a  rank-size rule fit was attempted with a  doubly decreasing power law in order to obtain an inflection point near the center of the data range, i.e. with the analytical form  \cite{ausloos}

\begin{equation}
y(r)\;=\;A \; m_1\; r^{-m_2}\;  (N-r+1)^{m_3},
 \label{Lav3} \end{equation}
 where $r$ is the rank, $A$ is an order of magnitude amplitude, {\it a-priori} imposed and adapted to the data, without loss of generality, for smoother convergence of the non-linear fit process, and $N$ is the number of regions, of course.
 The best  3-fit parameters, for $A = 10^3$ and $N=8092$,  have been so obtained:  $ m_1=0.847$;  $m_2= 0.68$;  $m_3= 0.209$: for a regression coefficient  $R^2= 0.957$, and a $\chi^ 2\ge$ 106 013, indicating a quite good agreement with the above equation (Fig. \ref {Plot20N20}). Some further discussion on some meaning of  the parameters $m_1$, $m_2$, and $m_3$ is postponed to Sect. \ref{Model}.

A similar fit study, made for the "regional ATI", is given in Fig.
\ref{fig:Plot35plots5yliloLavlike}, on a  semi-log plot for each
year.  The behavior being visually similar to that of Fig.  \ref
{Plot20N20} suggests to use Eq. (\ref{Lav3}) as well for further
study on  economic data.

Next,  we made an unweighted  average, over the quinquenium,  of  each  city ATI.  The rank-size relationship was looked for in  Fig.
\ref{Plot21ATI5yrav2fits}. The best fit parameters, for the function in Eq. (\ref{Lav3}),
for $A = 10^6$ and $N= 8092$, are $ m_1\sim27332$;  $m_2\sim0.938$;
$m_3\sim 1.05$: for a regression coefficient  $R^2= 0.985$, and a
$\chi^ 2\ge$ $10^{19}$, indicating a quite good agreement with the
above equation (Fig. \ref {Plot21ATI5yrav2fits}). However, the fit
is not very visually appealing.  It can be observed on a log-log
plot (Fig. \ref{Plot35plotslolo}) that  a few big cities (Roma,
Milano, Torino,  Genova, Napoli, Bologna,  Palermo and Firenze)
appear as outliers. We have removed these outliers from the overall
fits. When the top 8 cities are removed,  the best fit leads to   $
m_1\sim1.725$;  $m_2\sim0.725$;  $m_3\sim 0. 055$: for a regression
coefficient  $R^2= 0.998$, and a $\chi^ 2\ge$ $10^{15}$ (Fig.
\ref{Plot21ATI5yrav2fits}). The fit is better and more visually
appealing.

\subsection{Kendall $\tau$ coefficient}\label{Kendall coefficient}

The Kendall  $\tau$ measure is hereby discussed. Such a statistical
indicator compares the number of concordant pairs $p$ and
non-concordant pairs $q$,
i.e. how many times a  city occurs, or not, at the same ranks in both
(necessarily  equal size) lists.  This  measure is a usual correlation coefficient which allows to find whether the ranking of different measurements possesses some regularity.  In other words,
the Kendall  $\tau$ coefficient, measuring the cross correlation between two equal size data, is like  the cross-correlation function of two equal size time series \cite{kendall,Abdi}. The Kendall  $\tau$,  thus like the Pearson correlation coefficient,  allows a connection with  statistical physics theory:  in particular, it is an apparatus similar to the linear response theory correlation coefficients.
 For being more precise, notice that  $\tau$ is like the off-diagonal  generalized susceptibilities, in linear response theory \cite{kubo57,FTD_Kubo66_review},  in condensed matter \cite{Abdi}, since the variables are two different "fluctuations", an economic and a demographic one here.

By definition,
\begin{equation}\label{taueq}
\tau =  \frac{p-q}{p+q},
\end{equation}
thus suggesting how stable  the ranking is. 
Of course, $p+q= N(N-1)/2$, where $N$ is the
number of cities (8092 here), or the number of regions (20), in the
two (necessarily equal size) sets; thus, $p+q$ = 32 736 186 (cities)
or $p+q$=190 (regions).

For the computation of the Kendall $\tau$, i.e. to find $p$, $q$, and $p-q$, e.g.,  see \cite{wessa}, the procedure  in a stepwise form (the  case of cities is only outlined)  is the following:
\begin{itemize}
\item make a 2 column Table: the municipality name in column  (1) and the average ATI in column
(2);
\item do the same for the population data in column (4)
and  the municipality name in column  (3);
\item rank the cities in column (1) according to their average ATI, $r_{<ATI>}$, in column (2), for example in decreasing value order;
\item rank the cities in column (3) according to their  population size, $r_{Ninhab}$ in column (4), also in decreasing order;
\item  compare the position of cities ("ranked" columns (1) and (3)), i.e. find out how many  times    cities  occur  at the same  ranks in both ordering  (one obtains $p$) or at  different  ranks (for $q$).
\end{itemize}

Values of the Kendall   $\tau$,  Eq. (\ref{taueq}), $Z$,  Eq. (\ref{tauvar}), and other pertinent data for the correlations between the number of inhabitants, according to the 2011 census, and the average ATI over the quinquenium ($<2007-2011>$) are given in Table \ref{Tabletaurank}, as obtained from  Wessa   algorithm \cite{wessa}. 
 Observe from Table \ref{Tabletaurank} that $\tau \sim 0.85$.

From  a purely statistical perspective, under the null hypothesis of independence of the rank sets, the sampling would have an expected value $\tau = 0$. For large samples, it is common to use an
approximation to the normal distribution, with mean zero and variance, in order to emphasize the coefficient $\tau$ significance, through calculating:
 \begin{equation}\label{tauvar}
 Z=\frac{\tau}{\sigma_{\tau}}\;\equiv   \frac{\tau}{\sqrt{\frac{2(2N+5)}{9N(N-1)}}}.
     \end{equation}
Here, in the case of cities, $N=8092$ and $\sigma_{\tau} = 0.00741$.
Note that $\tau$ $\simeq 0.85$ ($\simeq 1$) and  $Z\simeq 115$.
Thus, it can be concluded that there is a strong regularity in the
pair ranking from a mere statistical point of view, -- even though
there are different regimes. In a thermodynamic sense, the system
presents different phases.

\subsection{Spearman's rank correlation coefficients}\label{Spearmancoefficient}


This section contains the computation of the Spearman $\rho$ and the
related discussion.

It is firstly needed to recall the definition of the Pearson
coefficient, as the ratio of the covariance of  two variables $x$
and $y$ to the product of their respective standard deviations, i.e.
\begin{equation}\label{Pearsonr}
\Pi= \frac{\Sigma xy - N\;(\Sigma x)(\Sigma y)}{\sqrt{[(\Sigma x)^2-N\;\Sigma x^2].[ (\Sigma y)^2-N\;\Sigma y^2]}}
\end{equation}in usual notations.

It is easy to show that the Pearson coefficient measures the
correlations between deviations form the mean,  i.e. correlations
between fluctuations, like the  transport coefficients in linear
response theory.  In the present case,  $\Pi$, like $\tau$,
corresponds to the off-diagonal  terms. Thus, it has  also some
direct statistical physics appeal.

The Spearman's  rank-order correlation coefficient $\rho$ is the
rank-based version of the Pearson correlation coefficient, i.e. the
values $x$ and $y$ of the measured quantities are replaced by their
corresponding rank in Eq. (\ref{Pearsonr}) (for computing the ranks,
see the first four bullets of the algorithm listed in the previous
Section):

 \begin{equation}\label{Spearmanrho}
\rho= \frac{\Sigma r_xr_y - N\;(\Sigma r_x)(\Sigma r_y)}{\sqrt{[(\Sigma r_x)^2-N\;\Sigma r_x^2].[ (\Sigma r_y)^2-N\;\Sigma r_y^2]}}   \equiv
\frac{\Sigma (r_x-<r_x>)(r_y-<r_y> )}{\sqrt{\Sigma (r_x-<r_x>)^2   \Sigma (r_y-<r_y>)^2}}
\end{equation}

It is worth noting that, except for the product of the rank
fluctuations appearing in the Spearman's form of Eq.
(\ref{Pearsonr}), the other terms are simply  related to the number
$N$ of measurements; e.g. $ \Sigma r_y= N(N+1)/2$.   In contrast,
Kendall $\tau$  reflects the number of concordances and discordances
regardless of the cardinality of the dataset, hence being a sort of
probability measure. Necessarily, Kendall's $\tau$ seems  to contain
more information on the distribution, and seems more reliable in
view of a statistical conclusion: indeed,  a few incorrect value
data  have less influence on the number of wrongly discordant pairs
than   the wrong absolute values would have on a Pearson, whence
Spearman also, coefficient, -- especially for finite size samples
\cite{BonnetWright}.

The Spearman coefficient has been calculated both at a municipal
level ($\rho \sim 0.9637$) and at a regional level ($\rho \sim
0.9098$), see Table \ref{Tabletaurank}.

High values have been found, as expected. In fact, usually Spearman
$\rho$ has a larger value than that of Kendall $\tau$, and in our
computations Kendall $\tau$ is rather large (see Table
\ref{Tabletaurank}). Therefore, Spearman $\rho$ confirms Kendall
$\tau$ outcomes on the regularity between economic and demographic
data both at a municipal as well as at a regional level.

 \begin{figure}
\centering
\includegraphics[height=10.0cm,width=12.7cm] {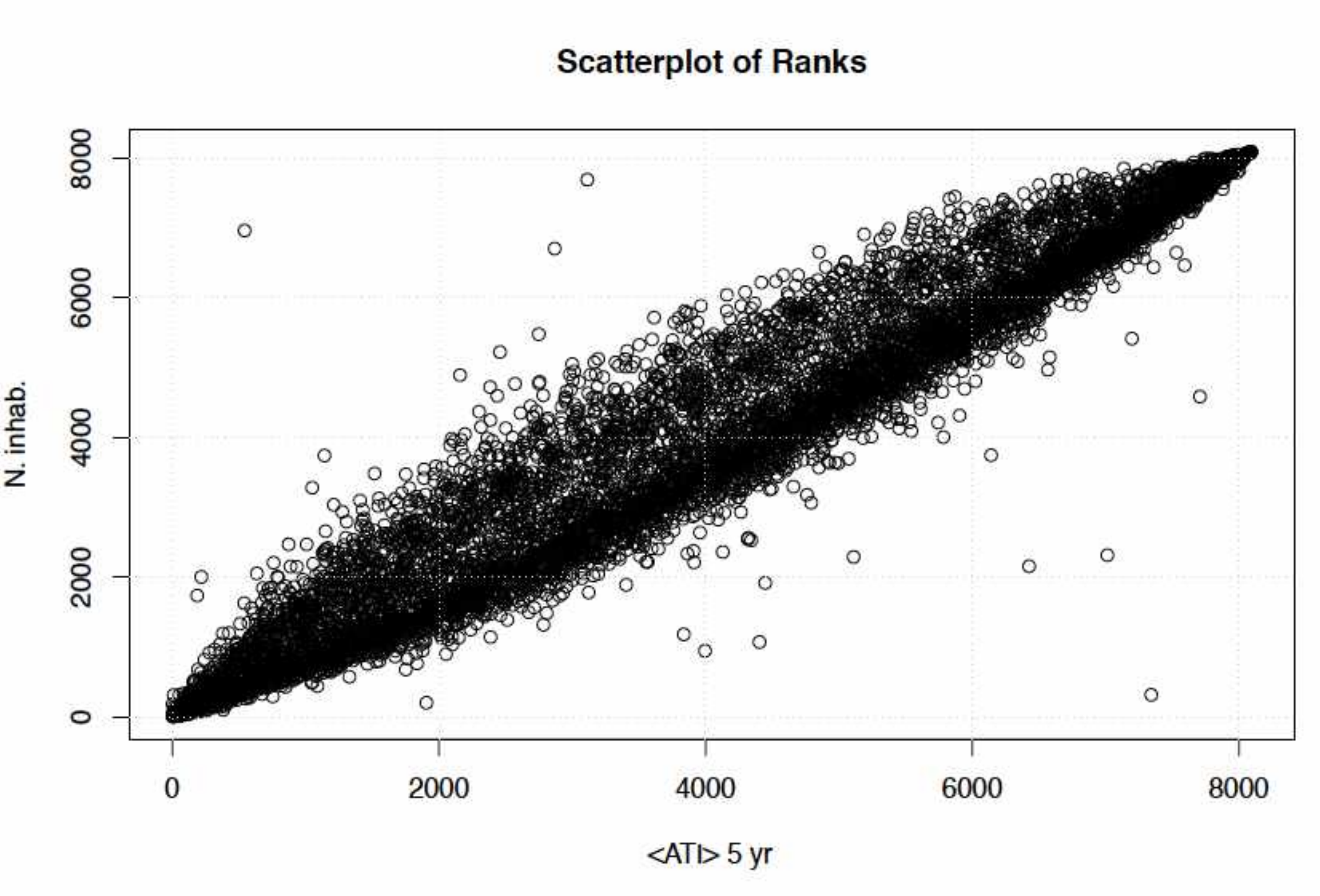}
\caption{ Scatter plot of the  city ranks for $<ATI>$  (averaged for
the  examined quinquennium)  and  the number of inhabitants in  each
Italian city.} \label{scatterplotrnksNATI}
\end{figure}

  \begin{figure}
\centering
\includegraphics[height=10.0cm,width=12.7cm] {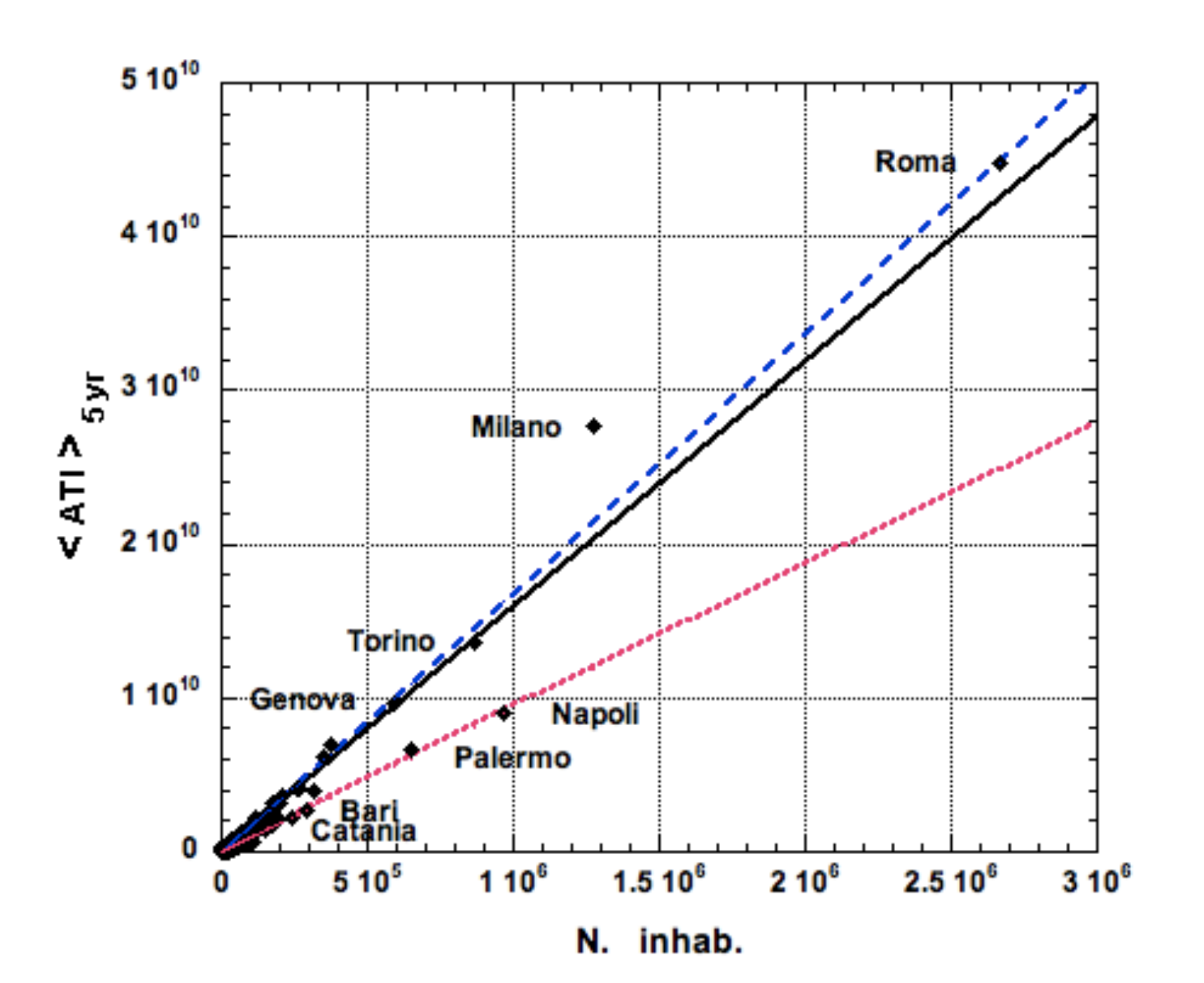}
\caption{ Scatter plot of the $<ATI>$  (averaged for the  examined
quinquennium)  and  the number of inhabitants in all Italian cities;
two sets of cities are emphasized from linear fits.}
\label{Plot8ATINscattplotlili}
\end{figure}

  \begin{figure}
\centering
\includegraphics[height=10.0cm,width=12.7cm] {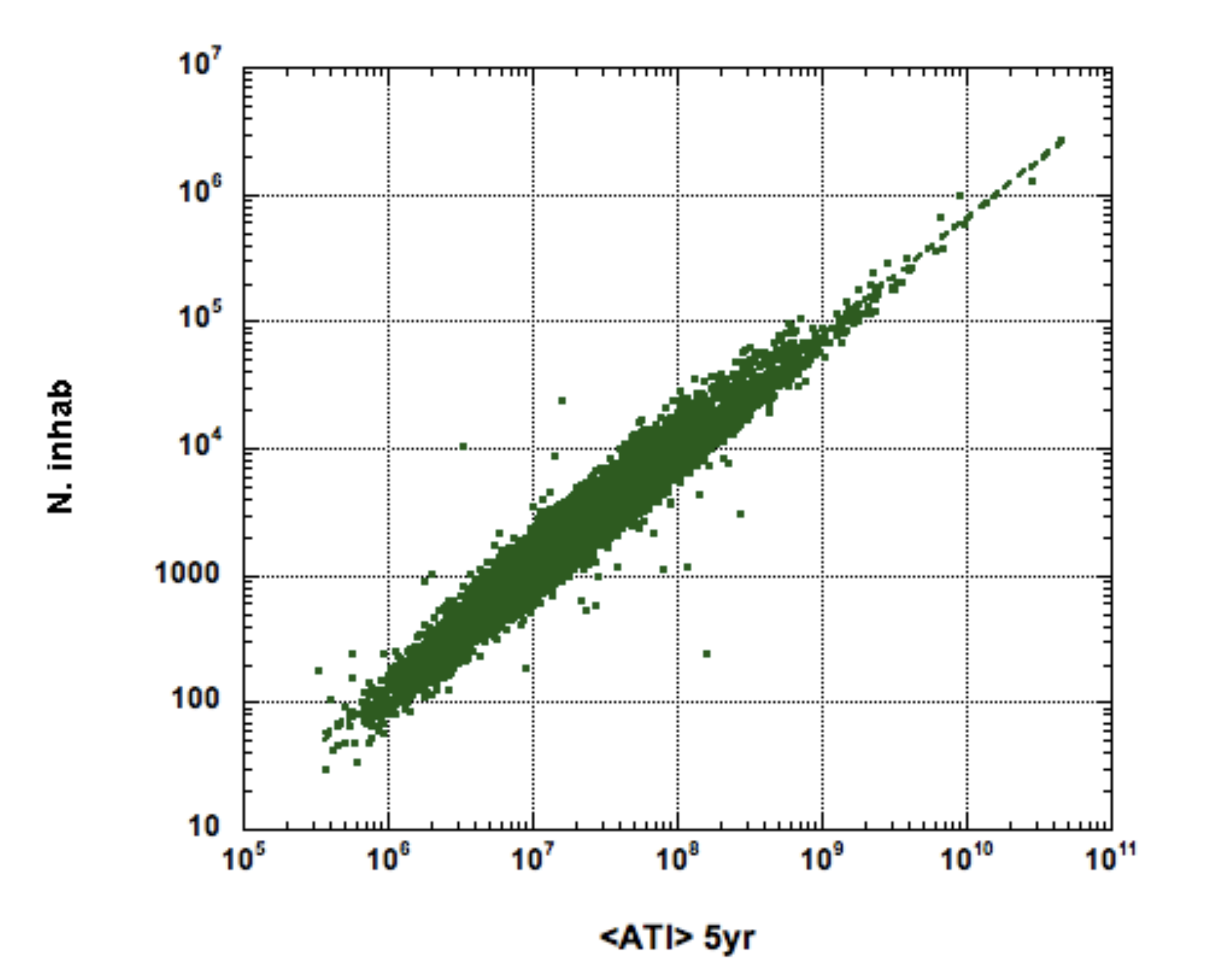}
 \caption{ Log-log scatter plot of  the
number of inhabitants and  the $<ATI>$  (averaged for the  examined
quinquennium)    in Italian cities; the main inertia axis is
indicated. } \label{Plot1NinhabATI5scat}
\end{figure}

 \begin{figure}
\centering
 \includegraphics[height=10.0cm,width=12.7cm] {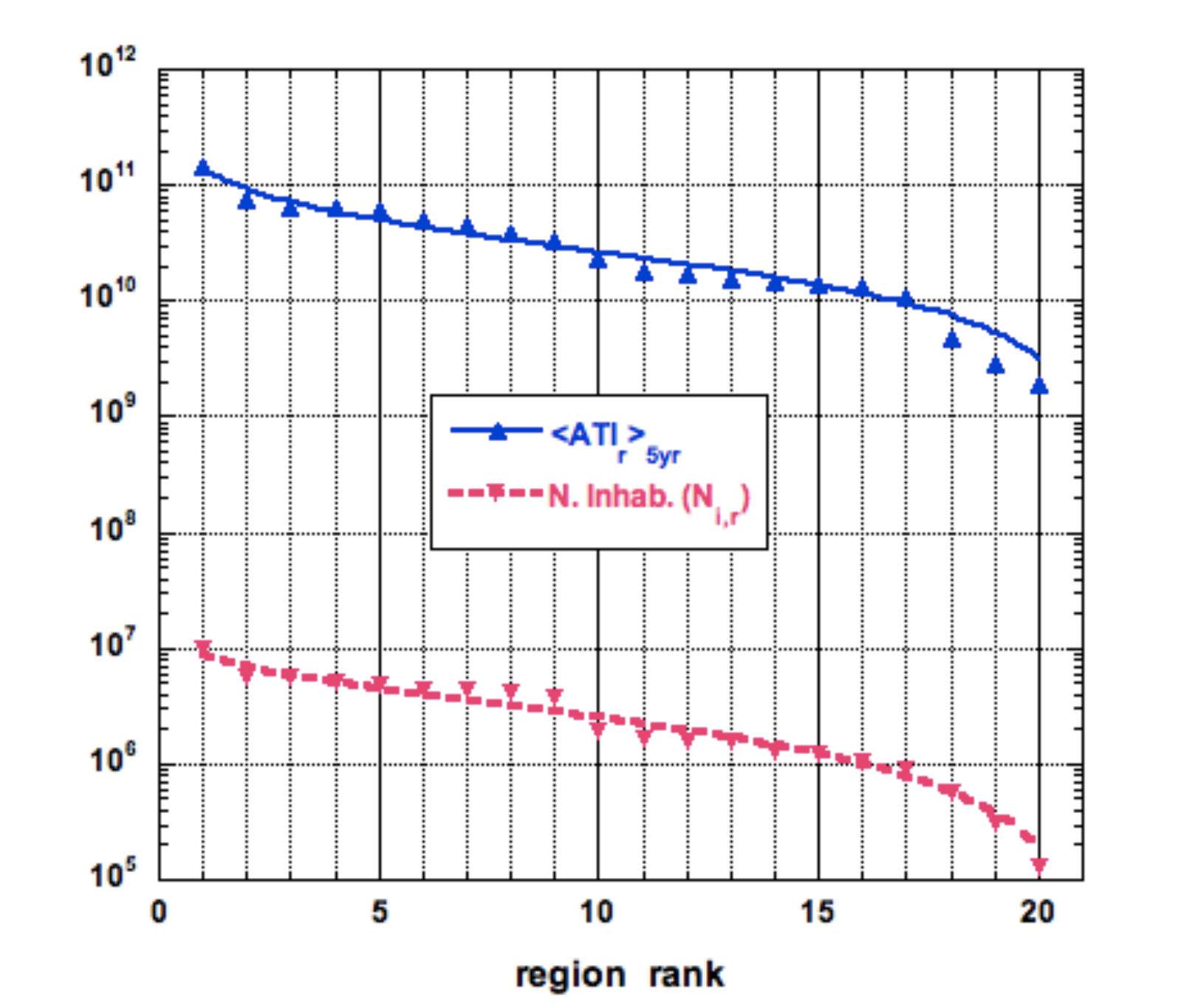}
\caption{ $N_{i,r}$ in IT  regions and averaged regional ATI $vs.$
the rank of the region for the years of the quinquennium. 
The fits correspond  to the function Eq. (\ref{Lav3});  the fit
parameters are given in the text.}
  \label{Plot4NavATILav3}
\end{figure}

   \begin{figure}
\centering
\includegraphics[height=10.0cm,width=12.7cm] {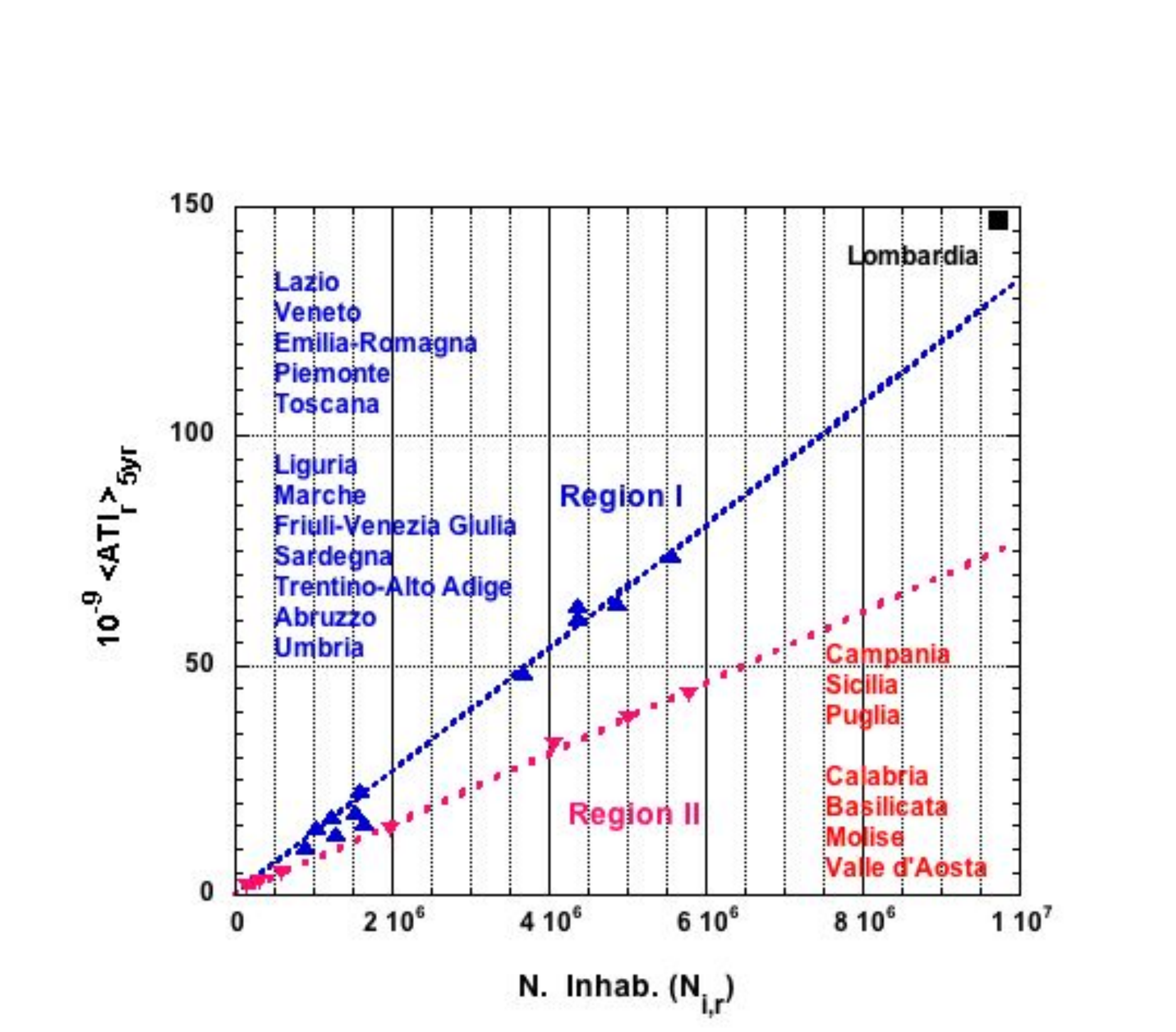}
\caption{ Scatter plot of the  region $<ATI_r>$ and the number of
inhabitants  ($N_{i,r}$) in Italian cities of the different regions;
two sets of regions  are emphasized from linear fits, - plus the
outlier Lombardia.}\label{Plot6NiATIscatterplotreg}
\end{figure}

 \begin{figure}
\centering
\includegraphics[height=10.0cm,width=12.7cm] {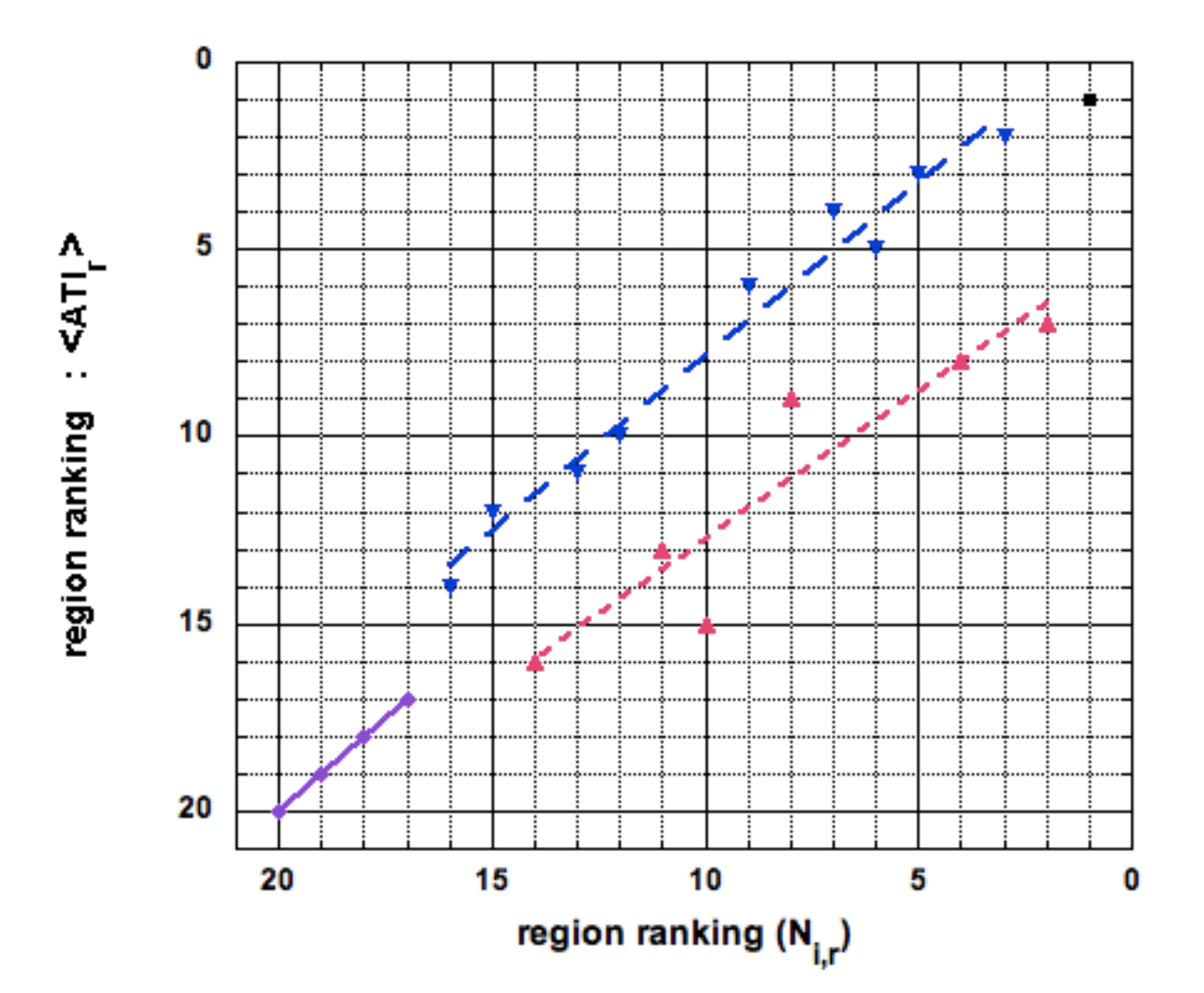}  
 \caption{ Scatter plot of the
region rank  for $<ATI>$ with respect to the    number of
inhabitants in a region rank; sets of regions  are emphasized by
linear best fits.} \label{Plot16scatterplotregrnks}
\end{figure}





\section{Results and discussion}\label{sec:results}

In this Section, the empirical  findings are commented upon  taking
into account numerical, economic, historical, demographic and
political considerations.

The scatter plot of cities rank-rank correlation (average ATI $vs.$
population) is shown in Fig. \ref{scatterplotrnksNATI}, as obtained
using \cite{wessa}. A large  number of cities are found to have
approximately the same rank,  within the elongated cloud of points,
but there are marked deviations. The main "inertia axis" can be
obtained: it reads: $r_{Ninhab} = 178.35(\pm15) + 0.956(\pm0.003)\;
r_{<ATI>}$. Some deviation from symmetry along the inertia axis is
observed.  A fine statistical analysis has shown  us that the
difference distribution $r_{Ninhab} - r_{<ATI>}$ is slightly
negative skewed; skewness $\sim -0.57$; the median = 92. Neglecting
the outlier tails, the distribution presents a smooth variation on
the negative  $r_{Ninhab} - r_{<ATI>}$ side, followed by  sharp peak
in the near 0 regime, itself followed by a sharp decay on the
$r_{Ninhab} - r_{<ATI>}$ range. This implies that  the probability
to find a  $r_{Ninhab} \le r_{<ATI>}$ is about 40\%. This suggests a
superposition of two homogeneous/similar distributions.

It is  also of interest to observe the scatter plot of the $<ATI>$
(averaged for the examined quinquennium)  and  the number of
inhabitants in all Italian cities; this is shown in Fig.
\ref{Plot8ATINscattplotlili}. Some structure inside the cloud of
data points can be emphasized:    two sets of cities  seem existing.
This is finely shown by visually distinguishing two sets of data
points, and  subsequent linear fits: one has (i) (blue dash line)
$y$ = 16 791.15  $x$, and (ii) (red dot line) $y$ = 9 311.28  $x$.
An overall fit gives  the proportionality (iii) (black continuous
line) $y$ = 15 942.30 $x$.  The fits cannot be compared through
their  regression coefficient; they are all close to 0.96, but  we
emphasize that the visual inspection  leads to some evidence. Notice
that  (i) Milano appears to be an outlier; (ii) the red dot line
seems to point to a set of cities "from the South"; (iii) in
contrast to the blue dash line, pointing cities "from the North".

The scatter plot of the number of inhabitants and the  $<ATI>$ is
also presented in Fig. \ref{Plot1NinhabATI5scat}, but  on a log-log
scale. This allows to emphasize the low values.  The two different
regimes are not well seen. It is like Fig.
\ref{Plot8ATINscattplotlili} with a change in $x$ and $y$ axes. A
power law fit through the cloud leads to the main inertia axis
equation given by $y  \simeq  0.456$ $10^{-3}$ $x^{0.915}$, with a
regression coefficient $R^2 = 0.963$.

Let us now again take two focussing points: (i) the cities in the
whole country and  (ii) the regions.

Recall that the regions having  a change over the quinquennium in
the number of cities are indicated by an arrow $\uparrow$ or
$\downarrow$;     the arrow direction is according to the change in
$N_{c,r}$ in some year as mentioned in Table
\ref{TableNcityperregion}. The fit in Fig. \ref{Plot20N20} captures
the administrative changes. It is based on the function in Eq.
(\ref{Lav3}); the fit parameters are given in the text.
Administrative changes are usually due to local tensions grounded on
historical motivations. Discussing such aspects is far beyond the
scopes of this paper. However, it is important to note that the
definition of the bounds of the IT regions derive often from the
administrative structure of Kingdoms and States in the Italian
territory after the Holy Roman Empire. In this respect, the
influence of the historical facts occurred in Italy (Napoleon, the
evolution of the Papacy, etc.) played also a relevant role.

Rank plots can be produced on classical, semi-log or log-log axes.
In the first case, the data looks like a mere decaying convex
function. However on semi-log (Fig.
\ref{fig:Plot35plots5yliloLavlike}) and on log-log (Fig.
\ref{Plot35plotslolo}), the ATI (and usually other data) shows some
structure.  An inflection point  is well seen near $r_M/2 \sim
4000$, on the 2007-2011 yearly ATI of the  8902 IT cities ranked
according to their "income tax" importance (Fig.
\ref{fig:Plot35plots5yliloLavlike}). Some jumps between $r=7$ and
$r=8$ are well marked on the log-log plot  (Fig.
\ref{Plot35plotslolo}). The rank plots are particularly meaningful
in describing the economical structure of Italy under the point of
view of the municipalities. The widest part of Italian cities has
comparable small amount of ATI;  this explains the inflection point
at $r_M \sim 4000$ and why the yearly ATI decreases vertically with
the rank for rank high enough. The jumps in the highest rank cities
identifies the great differences among the cities with highest
values of $<ATI>$. Such a difference is reduced for low  ranked
cities;  this leads to some understanding of  the polarization of
the aggregated (citizen)  income values in the main urban areas.

Such results are confirmed also by visually inspecting the semi-log
plot of the   rank-size  relationship between  each Italian city
$<ATI>$  (averaged for the  examined quinquennium) and its rank
(Fig. \ref{Plot21ATI5yrav2fits}). Some departure of the data from
the empirical fit  can be  noticed,   mainly  after the inflection
point. Specifically, this happens for the cities ranked above
$r\simeq 6000$, corresponding to an (averaged) ATI $\sim  10^7$ $�$.
These 2000 or so cities contribute to  $\sim 1.2 \;10^{10}$; mean
$\mu \sim  5.5\; 10^6$; standard deviation $\sigma \sim  2.6 \;10^6$
$�$. Thus $\mu/\sigma \sim 2.1$ for these cities. These cities
(roughly) correspond to those having less than 1000 inhabitants (the
border rank is at 6154.5), and $\mu \sim 543$, $\sigma \sim 256 $;
$\mu/\sigma \sim 2.1$ also. Such numbers are quite interesting,
mainly if one notes that the total IT population is about 5.957
$10^{7}$; $\mu \sim 7361$, $\sigma \sim 40262  $; $\mu/\sigma \sim
0.183$. Substantially, small cities have a number of inhabitants
which is, relatively to the mean,  less volatile when compared to
that of IT. This further confirms the polarization of Italian
inhabitants in a small number of highly populated cities.

The demographic, ATI relationship displayed through the scatter
plot of the  city ranks for $<ATI>$,   in  Fig.
\ref{scatterplotrnksNATI}, indicates a rather huge variation. However
the pair concordance is very high $\tau\simeq 0.849$ and $Z\simeq
114.6$.

Interesting  findings are seen in  Fig. \ref{Plot8ATINscattplotlili}, for the
scatter plot of the $<ATI>$  (averaged for the  examined
quinquennium)  and  the number of inhabitants in all Italian cities.
Two sets of cities are emphasized from linear fits. Such straight
lines capture a relevant aspect of Italian reality, which is divided
into different income distribution areas, the  South being much poorer
than the North. The red dot line includes cities showing a low
proportion between rank for $<ATI>$ and rank for population. In the
cities belonging to the blue dash line, such a proportion is high.
Cities in the former case (Bari, Catania, Palermo, Napoli) are
poorer than those of the latter one (Torino, Genova). Specifically,
Torino is less populated and richer than Napoli (and, similarly,
Genova is less populated and with a higher ATI than Palermo). The
reasons for this can be found in the well-documented distortion of GDP due to illegal activities and organized crime, which is more pervasive in the South than in the North
\cite{Brosio,Fiorio,Calderoni,Galbiati}.

The difference between the slopes of the red and blue
lines in Fig. \ref{scatterplotrnksNATI} may be useful in providing a
measure of the entity of shadow economy. Milano represents an
outlier for a simple reason: even if it is not the political capital
of Italy (it is Rome), it is the financial one (the Italian Stock
Market is in Milano). This explains the high value of ATI. Moreover,
Milano has a highly populated hinterland, with many big cities (like
Sesto San Giovanni or Rho). Hence, it is the center of a highly
populated area, even though the municipality of Milano itself  is not excessively
populated \textit{per se}, i.e. with respect to its ATI value.

\subsection{Regional disparities}\label{regionaldisparities}


Note for completeness, that the  number of provinces in 2007, i.e.
103, has increased by 7 units (BT, CI, FM, MB, OG, OT,
VS)\footnote{e. g. see ISO code:
\textit{http://en.wikipedia.org/wiki/Provinces$\_$of$\_$Italy.}} to
110 provinces in 2011. In this time window,  it is worth to point
out that 228  municipalities have changed from a province to another
one. Nevertheless, they remained in the same region, except for 7
cities from PU (the province of Pesaro and Urbino) in the Marche
region, to  RN (province of Rimini) in the Emilia Romagna region
(Casteldelci, Maiolo, Novafeltria, Pennabilli, San Leo, Sant' Agata
Feltria, Talamello). By looking at the data,  after  calculating
either the number ($N_{i,r}$) of inhabitants in a region or the
regional ATI ($ATI_r$), i.e. the sum for the relevant cities,  in
each year and the subsequent average, the change in regional
membership  appears to be very weakly relevant.



Thus, the regions can  be also  ranked each year according to their
$N_{i,r}$ and displayed on    a plot (Fig. \ref{Plot4NavATILav3}),
corresponding to $N_{c,r}$ in Fig. 1. Similarly, the regions can be
ranked according to   their yearly  $ATI_r$. For conciseness, this
is shown on this same Fig. \ref {Plot4NavATILav3}. The fit
parameters to Eq. (\ref{Lav3}),   with $A$ pre-imposed to be $=
10^9$,  are respectively  $m_1=0.445$ $10^ {-3}$; $m_2=0.287$;
$m_3=1.006$ with $R^2= 0.954$  for $N_{i,r}$, and $m_1=16.20$;
$m_2=0.54$; $m_3=0.719$, with $R^2= 0.966$ for  $<ATI_r>$.


It should not be necessary to repeat  that the rank of a region  is
not the same when ranking the $<ATI_r>$ (averaged for the  examined
quinquennium) and  when considering the number of inhabitants
$N_{i,r}$.  The comparison of  the region respective ranks is
however quite illuminating:  first, the Kendall $\tau$ calculation
can be easily performed; results are given in Table
\ref{Tabletaurank}, column (ii). The $\tau$ and $\rho$ values are
large ($\tau \sim 0.78$, $\rho \sim 0.9098$), but they are smaller
than when not distinguishing regions.

Remarkably, the data and fits (to the function in Eq. (\ref{Lav3}))
for $N_{i,r}$ in IT regions and averaged regional ATI $vs.$ the rank
of the region for the years of the quinquennium in Fig.
\ref{Plot4NavATILav3} indicate a coherence with respect to Fig. 1,
although  the data transformation is not that trivial. This shows
again that a rank-size rule is of great interest, showing structures
not seen when
  absolute value-size  relations are displayed or analyzed.

This is emphasized in the scatter plot of the  region $<ATI_r>$
(averaged for the  examined quinquennium)  and  the number of
inhabitants  ($N_{i,r}$) as well as the classical scatter plots in
Italian cities of the different regions;
  Fig. \ref{Plot6NiATIscatterplotreg} and Fig. \ref
{Plot16scatterplotregrnks}. Remarkably, it is visually found that IT
regions belong to different types of sets.  These sets of regions
can be emphasized also  through linear fits: (i)  a classical
scatter plot points to $three$ sets of regions, beside the outlier
(Lombardia). Furthermore,  the scatter plot of ranks indicate the
existence of subregions. Those sets are characterized by  a ratio
between the  ATI and the number of inhabitants,  either greater or
smaller than an "equilibrium point".

Fig. \ref{Plot6NiATIscatterplotreg} provides a regional confirmation
of the analysis carried out at the municipality level. The poor
regions are the Southern ones, while the cities in the North are
those belonging to the qualified group of high ATI. Valle d'Aosta
and Sardegna are peculiar cases of wrong classes (Valle d'Aosta is a
rich region belonging to the South group, for Sardegna the converse
applies). These findings appear to us not so meaningful, being Valle
d'Aosta and Sardegna positioned at the origin of the Cartesian plan
in Fig. \ref{Plot6NiATIscatterplotreg}. As for the cities,  Milano
is an outlier; for regions, Lombardia plays a similar  outlier role.
These outcomes describe well the situation of highly productive
regions in the North of Italy, with a South affected by the
organized crime and poor government institutions distorting
economical resources.

Also in this case, the gap between the  slopes of the blue and the
red lines may provide a good idea on how the ratio between
population and aggregated tax income should be (blue line, the
North), but  how it   is  presently in the  South (red line).

The rank-rank scatter plot of the region rank  for $<ATI>$ (averaged
for the  examined quinquennium)  with respect to the number of
inhabitants  in a region rank,  Fig. \ref{Plot16scatterplotregrnks},
is very interesting, and fits well with results obtained by the
reading of Fig. \ref{Plot6NiATIscatterplotreg}. Regions are
confirmed to be clustered in two groups (sets of regions are
emphasized by linear best fits). This is in  agreement with the
brief discussion here above on the Italian socio-economic
differences between the Northern regions and the Southern ones.

To conclude, it is worth noting the high variance, positive skewness
and kurtosis of the   distributions of ATI (see Table 2.).
Observe also the $\mu/\sigma$ values and time evolution in  this
Table 2 :  an increase in this variable indicates  a sort of
tendency to peaking of the sample distribution. Nevertheless, its
small value indicates   a quite  large variety in intrinsic ATI values
for the various cities. This effect is much obscured when looking at
the regional level, since then  $\mu/\sigma \simeq$ 1.12.


\section{Model}\label{Model}
A specific modeling is  presented based  on  arguments derived from   statistics.

First of all, it may be a wonder why a form like Eq. (\ref{Lav3}) is used. Observe that it can be related to a power law with exponential
cut-off \cite{Pwco3},  e.g.,  to the  Yule-Simon distribution   \cite{yule1922,BGNEGNKVZID}
  \begin{equation} \label{PWLwithcutoff}
 y(r)= h \;r^{-\alpha} \; e^{-\lambda r},
\end{equation}
appropriately describing settlement formation (following the
classical Yule model \cite{yule1922}) and its subsequent geographical distribution.

However, due to the  finite size of the number $N$ of data points, -
there cannot reasonably be an infinite amount of cities in a region,
the upper   $r$ regime should be considered as rather collapsing at
the  highest rank $r_M\equiv N$.   This  characterizes a function
with an  inflection point:  for  such a case, the Yule-Simon
distribution  can be adapted,    bearing upon the fact that $h\;
e^{-\lambda r} \equiv  \hat{d}\; e^{\lambda (r_M-r)}$, and
 \begin{equation}\label{asymptoticapprox}
e^{\lambda (r_M-r)} \simeq 1+ \lambda (r_M-r) \simeq  [1+(r_M-r) ]^{ \lambda }
\end{equation}
for $r \rightarrow r_M$,  thereby leading Eq. (\ref{PWLwithcutoff})
to  be written in the new form, that of  Eq. (\ref{Lav3}),
\begin{equation} \label{Lavalette3a}
 \;\; y(r)= \kappa_3\;  \frac{(N\;r)^{- \gamma}}  { (N-r+1)^{-\xi}  }
\end{equation}

The parameter $ \kappa_3$  (or $m_1$ in Eq. (\ref{Lav3})),  is like
the average  amplitude of the data, see $h$ in Eq.
(\ref{PWLwithcutoff}) also. Some meaning of the exponent $\gamma$ of
the  hyperbola, (or $m_2$ in Eq. (\ref{Lav3})),  can be obtained
from the decay exponent $\alpha$ in Eq. (\ref{PWLwithcutoff}).
Similarly, $\xi$ (or $m_3$ in Eq. (\ref{Lav3})), has  the meaning of
the decay exponent of an order parameter at a phase transition
\cite{Thompson,stanley,FisherRMP,StanleyRMP}

Usually, the parameters (exponents), like $m_2\leftrightarrow
\gamma$ and $m_3 \leftrightarrow \xi$, designate the statistical
physics model nowadays used for interpreting properties of  a
complex system, e.g. through phase transitions studies.

Having such ideas in mind, we suggest how to  interpret the
($m_2\leftrightarrow \gamma$ and $m_3 \leftrightarrow \xi$)
parameters through mathematical statistics theories, i.e. the
incomplete Beta function, 
 as follows.

Recall that a preferential attachment process is an urn process in
which additional balls   (e.g, settlement locations)  are added
continuously to the system and are distributed among the urns (e.g.,
areas)  as an increasing function of the number of balls the urns
already have. In the most general form of the process, balls are
added to the system at an overall rate of $m$ new species for each
new urn. This leads to the so called Yule-Simon probability
distribution
 \begin{equation} f(a;b) = b\,B(a; b+1).\end{equation} where $B (x; y) $ is the Euler Beta function
\begin{equation}B(x; y) = \frac{ \Gamma(x) \Gamma(y)}{\Gamma(x+y)},\end{equation}
  $\Gamma(x)$ being the standard Gamma function
  \cite{GradRyz,AbraSteg}.

In practical words, a newly created urn (= region) starts out with
$k_0$ balls (= cities) and further balls are added to urns at a rate
proportional to the number $k$ that they already have plus a
constant $a\ge -k_0$.  With these definitions,   the fraction $P(k)$
of urns (areas)  having $k$ balls (cities)  in the limit of long
time is given by
\begin{equation}
 P(k) = \frac{ B(k+a;b)}{B(k_0+a;b-1)} \end{equation} for $k\ge0$ (and zero otherwise).
In such a limit, the preferential attachment process generates a "long-tailed" distribution following  a  hyperbolic (Pareto) distribution, i.e.  power law,   
($\sim r^{-\alpha}$ or $\sim r^{-\gamma}$).

Moreover, a two-parameter generalization of the original Yule-Simon distribution replaces the Beta function with the incomplete Beta function:
\begin{equation}
B_{\epsilon} (a,b) = \int_0^\epsilon \;     x^a\;(1-x)^b\;dx
\end{equation}
  In statistics, the expression  $x^a(1-x)^b$ describes the probability of randomly selecting $a+b$ real numbers in [0,1] such that the first $a$ are in $[0,x]$ and the last $b$ are in $[x,1]$. The integral $\int^1_0 x^a\;(1-x)^b\;dx$
then describes the probability of randomly selecting $a+b+1$ real numbers such that the first number is $x$, the next $a$ numbers are in $[0,x]$, and the next $b$ numbers are in $[x,1]$.

It is worth noting that, in  most  of the studied examples concerning applications of the Beta-function in statistical physics, the number of urns   increases continuously, although this is not a necessary condition for a "preferential attachment".  In fact, it is unconceivable that an infinite number of urns regions) can be created.  Moreover,  an  increase in the number of settlements  (cities) is  limited by "available resources", e.g. by the socio-economic need for optimizing the useful distances between  settlements.

The product of two terms in Eq. (\ref{Lav3}) and the above reasoning remind of the   Verhulst's modification    \cite{Vlog3}  
  of the Keynes population expansion equation, when  introducing a  "capacity factor" . 

\section{Conclusions}\label{conclusions}

This paper   applies ideas of statistical mechanics in order to deal  with an  analysis of   cities and
regions. Specifically, the demographic (number of inhabitants, from Census 2011 - ISTAT) and the economic (ATI, averaged over the quinquennium 2007-2011, from MEF) ranking are compared and discussed for Italian cities.

Two statistical physics-like instruments have been mainly employed:
(i) a (new) rank-size rule  found as a doubly decreasing power law
type (see Eq.(\ref{Lav3}) and (ii) the computation of the Kendall
$\tau$ and Spearman $\rho$ coefficients for finding fluctuation
correlations, and phase state discrimination.
\begin{itemize}
\item
It is found that both cities and regions,  within the country, can be clustered in two  categories, which
mirror the Italian reality of a rich North and a poor South. Milano
(city) and Lombardia (region) represent outliers, and cannot be
indeed properly inserted in the resulting clusters.  Some social, economical and political arguments might be carried out to explain these findings. A few sentences have been introduced to suggest  reasoning outside the scope of this paper. It  has seemed  ppropriate to propose  a statistical physics-like model, based on  a number of evolving urn filling.
\item
Moreover, the above considerations and findings  also serve as a
demonstration of the advantage and  interest of the Kendall $\tau$
and Spearman $\rho$ coefficients to analyze and understand various
(equal size) lists  of variables measured according to various
criteria. It has been pointed out that such a measure is similar to
the fluctuation correlation coefficient in the linear response
theory of statistical physics. Interestingly, it provided an
indication of phase structures in the "sample" (=country).
\item For completeness, the Pearson $\Pi$ coefficient has been calculated. It has been argued that when the measurements are of so different natures, contain debatable error bars, rank-rank correlations are more meaningful, in contrast to the corresponding linear response theory coefficients in condensed matter physics.
\end{itemize}

It is worth saying, in concluding,  that the analysis at a provincial level might be of interest, but  this leads to complications  in the data and subsequent analysis. The impact of the creation of new provinces (103 $\rightarrow$ 110)  in the considered time period might  be interesting,   - such an administrative act,  similar to the application of an external field,   providing  an extra axis for investigations. 

\vskip0.5cm
 {\bf Acknowledgements}
 This paper is part of scientific activities in COST Action IS1104,
 "The EU in the new complex geography of economic systems: models,
 tools and policy evaluation".

\section*{Appendix A. On data reorganization}\label{5ITphago}

During the examined time interval, several cities have merged
into new ones, other were phagocytized.  For completeness and
thereby  explaining  some "finalized data reorganization",   we give
here below the various cases "of interest".  We use official IT
acronyms for the regions:

\begin{itemize}
\item[(i)]
Campolongo al Torre (UD) and Tapogliano (UD) have merged after a
public consultation, held on Novembre 27th, 2007, into Campolongo
Tapogliano (UD); thus 2 cities  $\rightarrow$ 1 city only;
\item[(ii)] Ledro (TN)
was the result of the merging (after a public consultation, held on
Novembre 30th, 2008) of
 Bezzecca (TN), Concei (TN), Molina di Ledro (TN), Pieve di Ledro (TN),
Tiarno di Sopra (TN) and Tiarno di Sotto (TN) as far as it is
explained e.g. in
$http://www.tuttitalia.it/trentino-alto-adige/18-concei/$; thus 6
$\rightarrow$ 1;
\item[(iii)] Comano Terme (TN) results from the merging of
Bleggio Inferiore (TN) and Lomaso (TN), in force of a regional law
of November 13th, 2009; thus 2 $\rightarrow$ 1;
\item[(iv)] Consiglio di Rumo (CO) and Germasino (CO) were annexed by
Gravedona (CO) on May 16th, 2011 and February 10th, 2011, to form
the new municipality of Gravedona ed Uniti (CO); thus 3
$\rightarrow$ 1.
\end{itemize}
To sum up:  13  cities (in 2007) $\rightarrow$ 4 cities (in 2011).
Thus,  the number 8092   taken as our reference number of   municipalities in the main text.

\section*{Appendix B. A short note on time dependence}\label{5yearKtau}
In this Appendix, it is verified that we do not bias the analysis
when we take the average of the ATI over the five year interval. In
order to verify the point, we have examined the ATI year-year
Kendall $\tau$ correlations  with respect to each other as well as
with respect to the  average\footnote{A Spearman $\rho$ has not been
   computed in such cases; it is not expected to provide
further insights. Indeed, Spearman $\rho$ is usually larger than
Kendall $\tau$;  $\tau$ is already remarkably large as seen  in Table
\ref{Tabletau}.}. From \cite{wessa}
 the $\tau$,  Eq. (\ref{taueq}),
and $Z$,  Eq. (\ref{tauvar}),
values  are easily obtained; the results are given in  Table
\ref{Tablepq} and Table \ref{Tabletau}. As mentioned in the main text, variations do exist but are rather mild.

Furthermore, a  bonus is  obtained in doing this time dependence examination. The
relevant quantities  given in Tables \ref{Tablepq}-\ref{Tabletau}
readily indicate a rather stable system of city hierarchies within
the examined time interval; e.g. $q/p \simeq 0.01$.

Scatter plots for every pair  and  for the scatter plots of pair of
ranks are available from the authors upon request.

\section*{Appendix C.  Pearson coefficient}\label{Pearson}
\begin{figure}
\centering
\includegraphics[height=10.0cm,width=12.7cm] {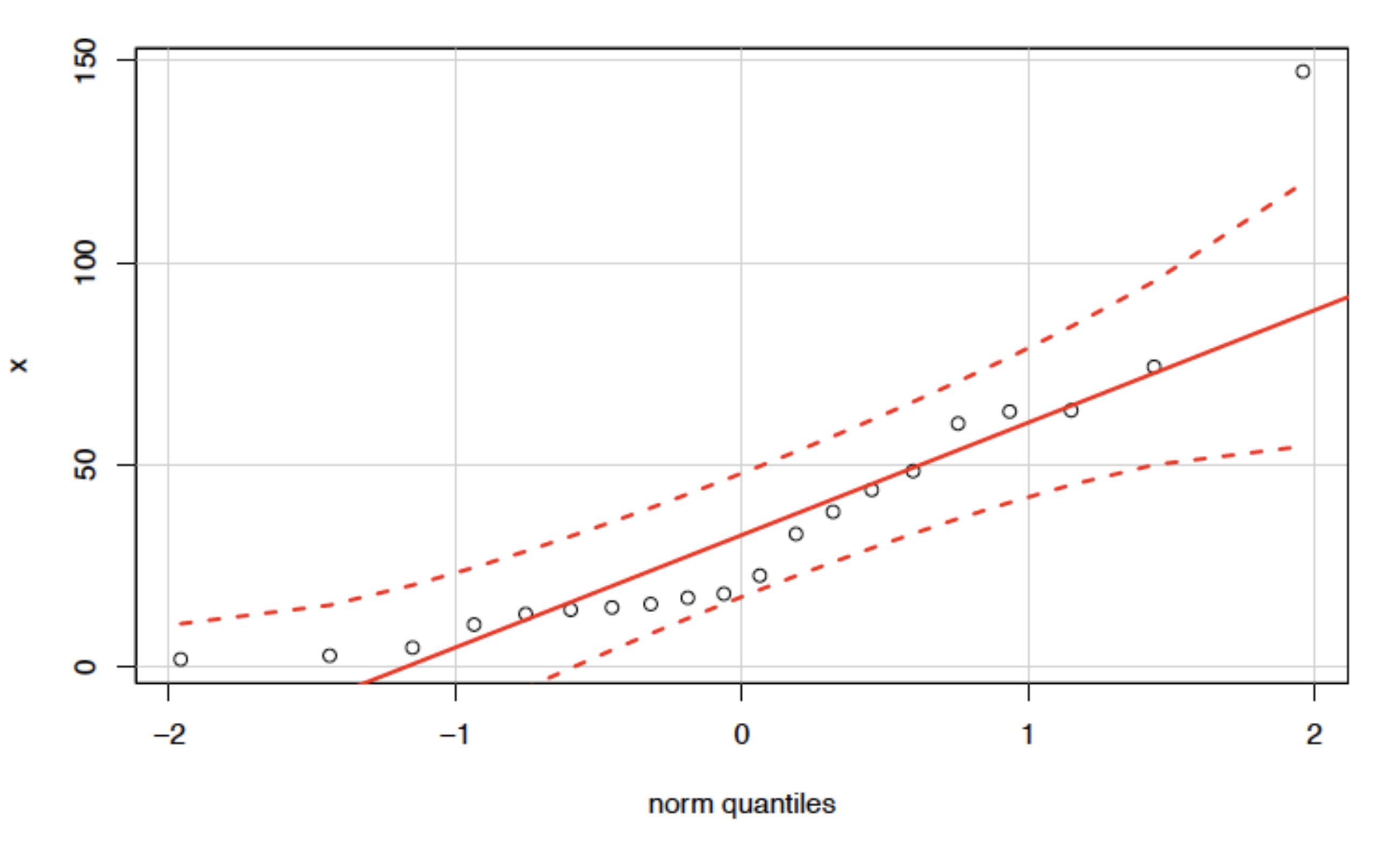}
 \caption{ Distribution of normalized quantiles for   $x$, i.e.,  the average ATI  for  cities throughout the 20 IT regions. } \label{QQx}
\end{figure}

\begin{figure}
\centering
\includegraphics[height=10.0cm,width=12.7cm] {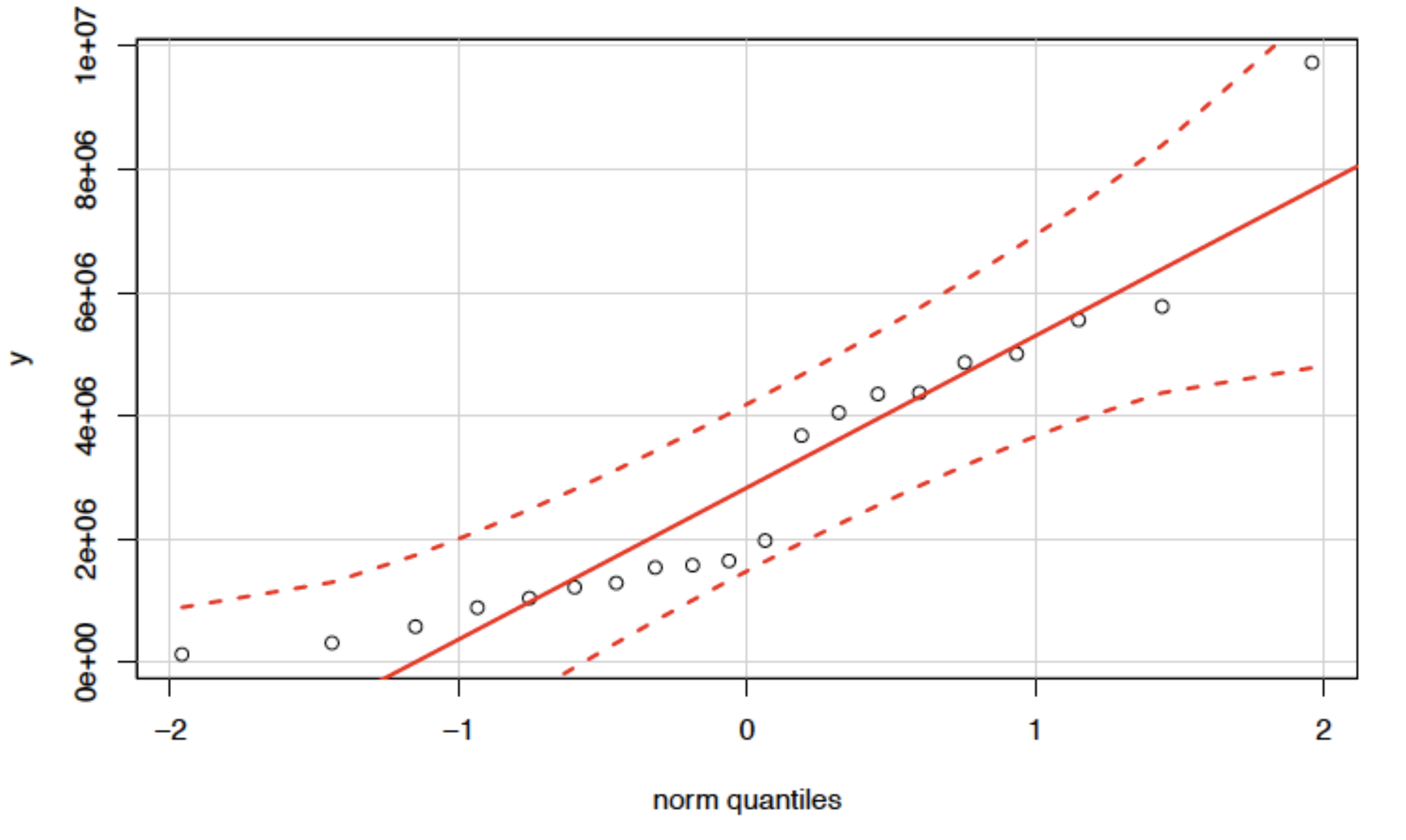}
 \caption{Distribution of normalized quantiles for  $y$, i.e. the number of  inhabitants in cities throughout the 20 IT regions. } \label{QQy}
\end{figure}

\begin{figure}
\centering
\includegraphics[height=10.0cm,width=12.7cm] {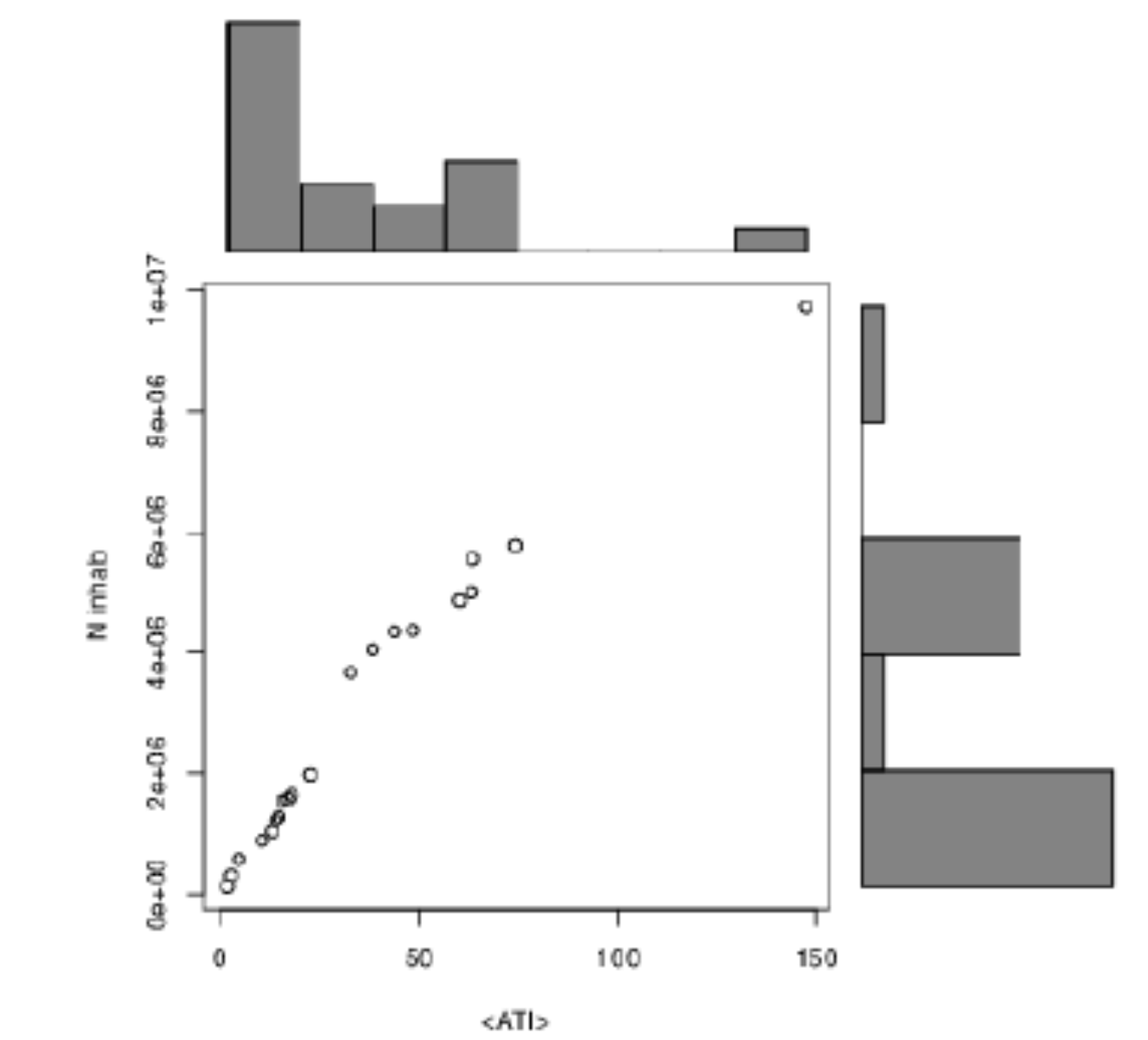}
 \caption{Relationship and distribution of the number of inhabitants and the average ATI  in  corresponding   IT regions. } \label{notnormallydistributed}
\end{figure}

The Pearson $\Pi$ coefficient, Eq.(\ref{Pearsonr}), is classically
used to estimate a correlation between (supposedly normally
distributed)  sets of measures  \cite{hays}.  The $\Pi$  values for
the case of the 8092 cities and the 20 regions, i.e. the value
correlations between the number of inhabitants (according to the
2011 Census) and the corresponding averaged ATI (over the period
2007-2011) are given in Table \ref{Tabletaurank}. Values are in the
same range as those of the Spearman $\rho$, but again much differ
from those of the Kendall $\tau$. It can be briefly argued that this
arises from the fact that the measurements are of different natures
and  found in intervals with  a quite  often unknown error bar, -
like many econo-sociological surveys: there  are not even similar
orders of magnitudes in measurements; there are outliers; the units
are wholly different ones.  Moreover, for calculating a  Pearson
$\Pi$  coefficient, and deducing its meaning, the measurements
should conform to some normality criterion; this is not quite the
case here.  Figures  \ref{QQx}-\ref{QQy}  and Fig.
\ref{notnormallydistributed} show that there are outliers and the
measurements  are not normality distributed.

In fact,  rankings are thought to be more illuminating and appealing, the more so when there is a sort of "competition". As a case surely already met by all readers, consider a hiring process in academia or a grant funding scheme to research groups: the hiring is strictly based on some ranking (and committee member consensus, of course or through some vote procedure); the funding is usually not based on the quantitative or qualitative values of  groups (they are usually  measured through different indicators, like for the case in our text),  relative  to each other. All conclusions mainly depend on some ranking correlations through  the chosen indicators. One ({\it in fine}) does not compare qualities. What is used is the rank. The �measured quality� has (alas) been by-passed. A Pearson correlation coefficient is rather irrelevant. The same is true in other "competitions", like in sport. The gap in points, the �quality� or �value�,  at the ned of a "season", is masked by the ranking, when one wants to glorify a team or an agent.  This goes also in many other cases. Recently,
     Raschke et al. \cite{PRERaschke}
have also concluded that the rank correlation is a more robust measure, in the  field of complex networks. This is exactly what we claim for the present case, but which is not {\it per se} a network.

\begin{table}  \begin{center}
 \begin{tabular}{|c|c|c|c|c| c|c|c|c|c| }\hline
 $q$ $\setminus$ $ p$   &   2007 & 2008&2009&2010& 2011&&$<5yav>$ \\\hline
2007&- &32322584 &32191636 &32162292 &32128014&&32294107\\
2008&413602&- &32476840 &32430158 &32381578 &&32561455\\
2009&544550 &259346&- &32544918 &  32466870&&32581689\\
2010&573894 &306028 &191268 &- &32530208&&32572191\\
2011&608172 &354608 &269316&205978 &-  &&32521347 \\ \hline
&&&&&&& \\\hline
$<5yav>$&442079 & 174731&154497&163995 &214839  &&-\\
   \hline
\end{tabular}
\caption{The number of concordant pairs  $p$   (above the diagonal)
and that of non-concordant pairs $q$ (below the diagonal)  of the
8092 cities, according to their ATI value, for every year
pairs.}\label{Tablepq}
 \end{center}
 \end{table}

\begin{table}  \begin{center}
 \begin{tabular}{|c|c|c|c|c| c|c|c|c|c| }\hline
$Z$ $\setminus$ $\tau$&   2007 & 2008&2009&2010& 2011&&$<5yav>$ \\\hline
2007&- &0.9747 &0.9667 &  0.9649 &0.9628& &0.9730\\
2008&131.49&- &0.9842 &0.9813 &0.9783 &&0.9893\\
2009&130.42  &132.77&- &    0.9883  &0.9835& &0.9906\\
2010&130.18 &132.38&133.32 &-  &    0.9874 & &0.9900\\
2011&129.89 &  131.98 &132.68&133.21 &-  &&0.9869\\   \hline
&&&&&&& \\\hline
$<5yav>$&131.26 &  133.47 &133.63& 133.55& 133.13& &-\\
   \hline
  \end{tabular}
\caption{ Rank correlation test of the 8092 cities, according to
their ATI value: $\tau$ (above the diagonal), Eq. (\ref{taueq}), and
$Z$ (below the diagonal), Eq. (\ref{tauvar}), for every year pairs.}
\label{Tabletau}
 \end{center}
 \end{table}

\end{document}